\documentclass{elsart}
\usepackage{graphicx}
\usepackage{amssymb}
\newcommand{\be}{\begin{equation}}
\newcommand{\ee}{\end{equation}}
\newcommand{\bea}{\begin{eqnarray}}
\newcommand{\eea}{\end{eqnarray}}
\newcommand{\bvec}[1]{\mbox{\boldmath $#1$}}
\newcommand{\eqref}[1]{(\ref{#1})}
\newcommand{\F}{\phantom {1}}

\begin{document}
    
\begin{frontmatter}

\title{Static potential, force, and flux-tube profile
in 4D compact U(1) lattice gauge theory 
with the multi-level algorithm}

\author{Yoshiaki Koma}, \ead{ykoma@mppmu.mpg.de}
\author{Miho Koma}, \ead{mkoma@mppmu.mpg.de}
\author{Pushan Majumdar}\footnote{present 
address:
Institut f\"ur Theoretische Physik,
Karl-Franzens-Universit\"at Graz, Austria
} 
\ead{pushan@mppmu.mpg.de}

\address{Max-Planck-Institut f\"ur Physik,
F\"ohringer Ring 6, D-80805, M\"unchen, Germany}

\begin{abstract}
The long range properties of four-dimensional 
compact U(1) lattice gauge theory with the Wilson action 
in the confinement phase is studied 
by using the multi-level algorithm.
The static potential, force and flux-tube profile    
between two static charges
are successfully  measured from the correlation function
involving the Polyakov loop.
The universality of the coefficient of the $1/r$ 
correction to the static potential, known as the 
L\"uscher term, and the transversal width of the flux-tube 
profile as a function of its length are investigated.
While the result supports the presence of the 
$1/r$ correction, the width of the flux tube 
shows an almost constant behavior at a large distance.
\end{abstract}

\end{frontmatter}

\section{Introduction}
\label{sec:sec1}

\par
There is a conjecture that (non supersymmetric) confining
gauge theories in the infrared regime are
related to an effective bosonic string theory.
If this happens,  the asymptotic behavior of the potential
between static charges separated by distance $r$
is expected to be parametrized as
\be
V(r) = \sigma r + \mu + \frac{\gamma}{r}+O(\frac{1}{r^{2}}).
\label{eqn:asympt-pot}
\ee
Here $\sigma$ is the string tension, which characterizes
the strength of the confining force of static charges, and
$\mu$ denotes a constant.
The third term is the zero point Casimir energy for an open 
bosonic string with fixed boundary. 
This correction is known as the 
L\"uscher term~\cite{Luscher:1980fr} and the coefficient
$\gamma$ is considered to be universal,
such that it does not depend on the gauge group
but only on the space-time dimension $d$
through $\gamma = -\pi (d-2)/24$.
The effective bosonic string theory also
predicts that the width of the field energy distribution
of the flux tube  diverges logarithmically as 
$r \to \infty$~\cite{Luscher:1981iy}.
Recent Monte Carlo simulations of various
lattice gauge theories support the universality of
$\gamma$ in Eq.~\eqref{eqn:asympt-pot} 
with high accuracy: 
the confinement phase of 
$Z\!\!\!Z_{2}$ lattice gauge theory (LGT)  in 
3D~\cite{Caselle:1997ii,Caselle:2002rm,Caselle:2002ah},
SU(2) LGT in 3D~\cite{Majumdar:2002mr,Kratochvila:2003zj} and in 
4D~\cite{Bali:1995de},
SU(3) LGT in 3D~\cite{Luscher:2002qv} 
and in 4D~\cite{Luscher:2002qv,Juge:2002br}.
Moreover, in Refs.~\cite{Majumdar:2002mr,Juge:2002br}, the 
spectrum of the string states have  
been computed, which further support the effective 
string description of confining gauge theories.

\par
In this context, we are now interested in the 4D 
compact U(1) LGT with the Wilson action.
This theory possesses a confinement phase 
analogous to non-Abelian gauge theories
below the critical coupling 
$\beta < \beta_{c}\approx 1.011$
(precise value can be found in Ref.~\cite{Arnold:2000hf}).
This is due to the presence of magnetic 
monopoles~\cite{DeGrand:1980eq},
which cause the dual Meissner effect
like in a dual superconductor~\cite{Banks:1977cc,Peskin:1978kp}
when electric sources are introduced in the vacuum;
the electric flux is squeezed into a flux tube by
the circulating monopole supercurrent, which leads to
a linear rising potential between static charges.
In fact, the measurements of the U(1) flux-tube profile 
in the confinement phase have been reported
in Refs.~\cite{Singh:1993vc,Zach:1995ni}, which 
support the dual superconductor picture.
To answer the questions 
i) whether the static potential in this theory also 
contain the universal correction,
ii) how the width of the flux-tube profile behaves as a 
function of $r$,
in this paper, we investigate the long range properties of the
potential, force 
and the flux-tube profile between two static charges.
Here we are going to use
the Polyakov loop correlation function 
(PLCF: a pair of Polyakov loops separated by a distance $r$)
as an external source.
Contrary to the use of the Wilson loop $\langle W(r,t)\rangle$, 
if we use the PLCF $\langle P^{*}P(r) \rangle$, 
we do not need to care about the 
$t$ dependence of the results and can extract the
the ground state easily as long as the lattice temporal extension 
is large enough.
This is important because most of analytical predictions 
are  given for such a ground state.
However, the measurement of the PLCF 
in the confinement phase with large $r$
is a quite difficult task
since the expectation value becomes 
exponentially smaller with increasing $r$.
Moreover, due to the strong coupling nature of the theory,
the signal-to-noise ratio is very small from the beginning.
In principle, one would need enormous statistics and 
computation time to identify such small expectation values.

\par
Recently, L\"uscher and Weisz (LW) have proposed 
a multi-level algorithm for pure LGT~\cite{Luscher:2001up}
to compute the expectation value of
a Wilson loop for a large size and a
PLCF for large $r$ with exponentially
reduced statistical errors.
They noted that its algorithmic idea is essentially 
the same as in the multi-hit method~\cite{Parisi:1983hm} 
but is applied to pairs of links instead of single links.
In fact,  based on their algorithm they have
confirmed the presence of an universal 
potential $\gamma/r$ in SU(3) LGT~\cite{Luscher:2002qv}.
The LW algorithm is applicable to other LGTs 
as long as a local gauge action is simulated.
The Wilson action is the easiest gauge action to adopt.
The studies of 3D SU(2) 
LGT~\cite{Majumdar:2002mr,Kratochvila:2003zj} 
also use this algorithm.
We then expect that the LW algorithm 
can also be applied to compact U(1) LGT, which
will help to overcome its numerical difficulties.
In the original work~\cite{Luscher:2002qv}, although 
the potential and its derivatives with respect
to $r$, force etc., have been of interest,
we find that this is also applicable to 
measuring the flux-tube profile
as well as the glueball mass 
measurements~\cite{Majumdar:2003xm}.

\par
In section~\ref{sec:sec2}, we describe  
the LW algorithm for the measurements of the static 
potential and force from the PLCF.
We also explain how to measure the 
flux-tube profile in this context.
In sections~\ref{sec:sec3} and~\ref{sec:sec4}, 
we present simulation details and the numerical results
on the PLCF/potential/force and on the the flux-tube profiles,
respectively, where several analyses of the data are given.
The section~\ref{sec:sec5} is a summary.
A part of these studies has been presented at the
Lattice 2003 at Tsukuba, Japan~\cite{Koma:2003gv}.

\section{Numerical procedures}
\label{sec:sec2}

\par
In this section we describe how to measure
the static potential, force and flux-tube profile 
with the LW algorithm.

\par
The Wilson gauge action of the compact U(1) LGT is given by
\be
S[U] = \beta \sum_{m} \sum_{\mu <\nu}
\left \{ 1- {\rm Re}\; [U_{\mu\nu}(m)] \right \} \; ,
\ee
where $U_{\mu\nu}(m) \in U(1)$ are  plaquette variables  
constructed from link variables
$U_{\mu}(m)=\exp (i\theta_{\mu}(m)) \in U(1)$.
We consider the lattice 
volume $N_{s}^{3} \times N_{t}$,
where $N_{t}$ is an even number.
The four-dimensional sites are labeled by
$m=(m_{s},m_{t})$ with $m_{s}=(m_{1},m_{2},m_{3})$.
Periodic boundary conditions are imposed for all directions.
A sequence of independent gauge field configurations, 
labeled by $i_{c}=1,2,\ldots,N_{c}$, 
are generated by 
using a mixture of heatbath (HB) and over-relaxation (OR)
link updates as usual.

\subsection{The static potential and force from the PLCF}

\par
We adopt the terminology of Ref.~\cite{Luscher:2001up}.
To measure the PLCF $\langle P^{*}P(R) \rangle$
for a charge distance $R=r/a$,
we first take sub-lattice 
averages of the two-link correlators
\be
\mathbb{T}(m;R;i)= U_{4}^{*}(m) U_{4}(m+ R \; \hat{i}) \; ,
\label{eq:two-link1}
\ee
as
\be
\mathbb{T}^{(2)}(m_{s},\bar{m}_{t}; R;i)
= [\mathbb{T}(m ; R;i)\mathbb{T}(m+\hat{4};R;i)] \; ,
\label{eq:two-link2}
\ee
where $i=1,2,3$ are possible directions of two static charges
and $\bar{m}_{t}=1, 3, \ldots, N_{t}-1$.
The sub-lattice average is achieved 
by updating link variables (with a mixture of HB/OR)
except for the spatial links at the time slice $\bar{m}_{t}$.
We call this procedure the {\em internal update}.
We repeat the internal update until 
reasonably stable values for $\mathbb{T}^{(2)}$
are obtained.

\par
\begin{figure}[!t]
\centering
\includegraphics[height=4.1cm]{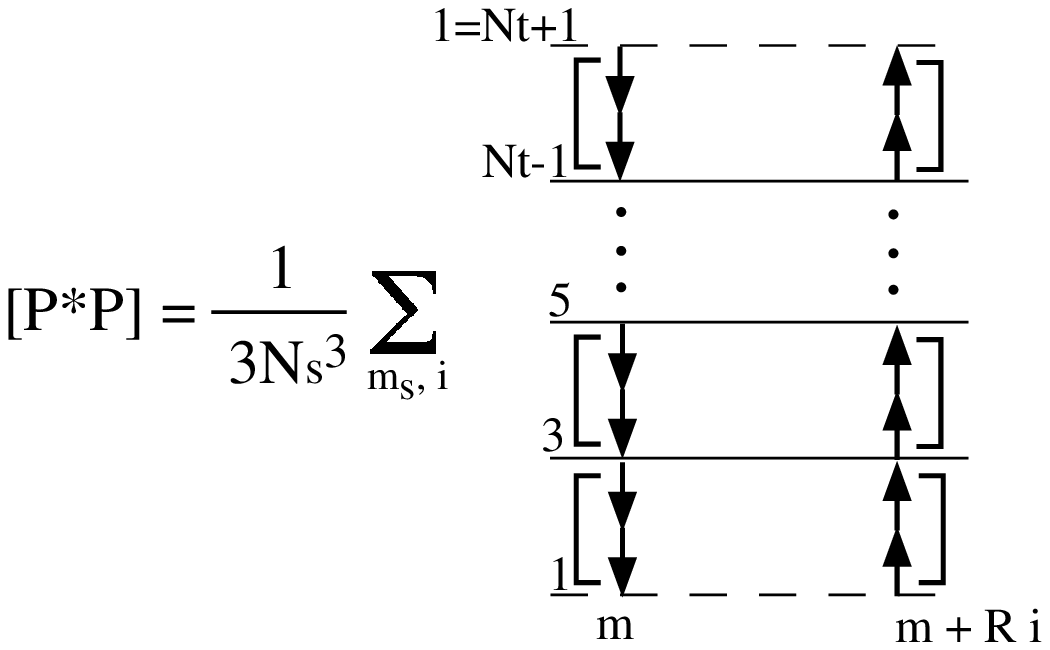}
\caption{How to construct the $[P^{*}P]_{i_{c}}$
with the LW algorithm. 
$[\cdots ]$ denotes
the sub-lattice average.}
\label{fig:LW-poly}
\end{figure}

Then, the PLCF at a spatial site 
$m_{s}$  is constructed from  $\mathbb{T}^{(2)}$ as
\bea
&&
P^{*} P (m_{s};R;i)
=
P^{*} (m_{s}) P (m_{s}+ R\hat{i}) \nonumber\\*
&=&
{\rm Re} \; 
\mathbb{T}^{(2)}(m_{s},1;R;i) \mathbb{T}^{(2)}(m_{s}, 3; R; i)
\cdots
\mathbb{T}^{(2)}(m_{s}, N_{t}-1; R; i) \; .
\label{eqn:poly1}
\eea
We take the average 
with respect to $m_{s}$ and $i$, which provides
the value of the PLCF  for $i_{c}$th
configuration, $\left [ P^{*} P (R) \right ]_{i_{c}}$.
For a schematic understanding, see Fig.~\ref{fig:LW-poly}.
The desired expectation value 
$\langle P^{*} P (R) \rangle$ is calculated from the average of 
$\left [ P^{*} P (R) \right ]_{i_{c}}$ for $i_{c}=1,2,\ldots,N_{c}$.

\par
The static potential and the corresponding force 
are taken as (neglecting terms of $O(e^{-(\Delta E) N_{t}})$)
\bea
aV(R) 
&=&  -\frac{1}{N_{t}} \ln  \langle P^{*}P (R) \rangle \; ,
\label{eqn:potential}\\
a^{2} F(\bar{R}) &=& aV(R)-aV(R-1)  \; ,
\label{eqn:force}
\eea
where $\bar{R}=R-1/2$.

\par
In the actual measurements, 
we have also applied the multi-hit technique~\cite{Parisi:1983hm} 
to the timelike link variables $U_{4}(m)$ for $R \ge 2$
before constructing the two-link correlators~\eqref{eq:two-link1}.
It was further helpful to reduce the statistical errors
of the PLCF within a limited CPU time.
In U(1) LGT, this procedure is given by the replacement
\be
U_{\mu}(m) \mapsto 
\frac{I _{1}(\beta |W_{\mu}(m)|)}
{I_{0}(\beta |W_{\mu}(m)|)} 
\frac{W_{\mu}(m)}{|W_{\mu}(m)|} \; ,
\ee
where $I_{0}$ and $I_{1}$ denote the zeroth- and  first-order
modified Bessel functions and
$W_{\mu}(m)$ the sum of six staples around $U_{\mu}(m)$:
\bea
\!\!\!\!\!\!\!\!\!\!
W_{\mu}(m) 
\! =  \!
\sum_{\nu \neq \mu}
  \Bigl[
  U_{\nu}^{*}(m \! + \! \hat{\mu})
  U_{\mu}(m \! +\! \hat{\nu})U_{\nu}(m)
\! + \!
  U_{\nu}(m \! + \! \hat{\mu} \! 
  - \! \hat{\nu})U_{\mu}(m \!-\! \hat{\nu})
  U_{\nu}^{*}(m)
  \Bigr] \; .
\eea

\par
Some comments associated with the LW algorithm are in order.
The number of the internal updates $N_{\rm iupd}$ is
an optimization parameter that has to be
tuned for efficient performance of the computation.
If one wants to compute the force, 
which requires two values of the potential 
at different $r$,
it is useful to compute all $R=1,2,\ldots, R_{\rm max}$ 
in one run without changing $N_{\rm iupd}$ depending 
on $R$, although a small number of $N_{\rm iupd}$ is enough
for a short distance.
This is because data among different $R$'s
are highly correlated, which leads to
a significant cancellation of the statistical 
errors in the difference~\cite{Luscher:2002qv}.
Practically, one may regard not only the 
PLCF but also the potential and force as the primary observables 
and apply the jackknife analysis for the evaluation of the
statistical errors.

\subsection{The flux-tube profile}
\label{subsec:procedure2}

In order to measure the flux-tube profile, one needs
to compute a correlation function of the type
\bea
\langle  \mathcal{O} (n) \rangle_{j} =
\frac{\langle  P^{*}P \mathcal{O}(n) \rangle_{0}}
{\langle P^{*}P \rangle_{0}} 
- \langle \mathcal{O}\rangle_{0} \; ,
\label{eq:profile-correlator}
\eea
where $\mathcal{O}(n)$ is a local operator, 
$\langle \cdots \rangle_{j}$ denotes an average 
in the vacuum with the PLCF,
and $\langle \cdots \rangle_{0}$ an average in the vacuum 
without such a source.
For a parity-odd local operator, we do not need 
the second term since it gives no 
contribution, $\langle \mathcal{O}\rangle_{0}=0$.
However, if one is interested in a parity-even local operator
such as the action density $\cos \theta_{\mu\nu}(n)$,
where $\theta_{\mu\nu}(n)$ is the phase of $U_{\mu\nu}(n)$,
one needs to subtract out the vacuum expectation value.
It is noted that to receive  maximum benefit from the LW 
algorithm, the parity-odd local operator
is preferable~\cite{Majumdar:2003xm}.

\par
\begin{figure}[t]
\centering
\includegraphics[height=4.1cm]{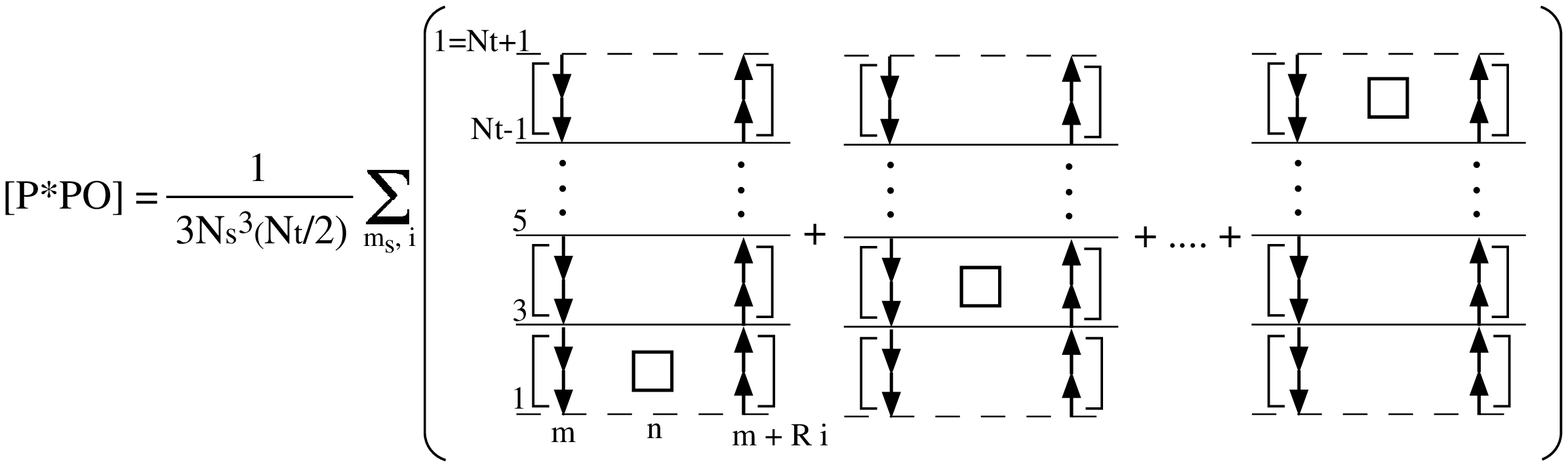}
\caption{How to construct $[P^{*}P\mathcal{O}]_{i_{c}}$
with the LW algorithm. 
$[\cdots]$ denotes the sub-lattice average
 and the square represents a local operator.}
\label{fig:LW-poly-ope}
\end{figure}

To measure $\langle P^{*}P \mathcal{O} (n)\rangle_{0}$ 
on the mid-plane between two static charges, 
we parameterize the position of the local operator $n$ as
\be
n=m+(R/2) \hat{i} + x \hat{j}+y\hat{k} \; ,
\ee
where $i$ is the direction of two static charges and
$j-k$ specify a 2D plane perpendicular to $i$.
By constructing the two-link-local-operator correlators
\be
\mathbb{O}(m; n ; R; i) 
= 
U_{4}^{*}(m) U_{4}(m +R \hat{i}) \mathcal{O}(n) \; ,
\ee
we compute sub-lattice averages of the correlation function
\be
\mathbb{TO}^{(2)}(m_{s},\bar{m}_{t} ; n ; R ; i) 
= 
[\mathbb{T}(m; R ;i )\mathbb{O}(m+\hat{4}; n; R; i) ]\; .
\label{eqn:ope-te}
\ee
Combining $\mathbb{TO}^{(2)}$ and 
$\mathbb{T}^{(2)}$ in Eq.~\eqref{eq:two-link2},
we obtain the PLCF involving a local operator at
site $m_{s}$ as
\bea
&&
\!\!\!\!\!\!\!\!\!\!
P^{*} P  \mathcal{O} (m_{s} ; x, y ; R ; i) \nonumber\\
&&
\!\!\!\!\!\!\!\!\!\!
= \frac{1}{(N_{t}/2)}
{\rm Re} \;
\Biggl \{ 
\mathbb{TO}^{(2)}(m_{s}, 1 ; n; R;i) 
\mathbb{T}^{(2)}(m_{s},   3 ;R ;i)
\cdots
\mathbb{T}^{(2)}(m_{s},  N_{t}-1; R;i) +
\nonumber\\
&&
\cdots +
\mathbb{T}^{(2)}(m_{s}, 1 ; R;i) 
\mathbb{T}^{(2)}(m_{s},   3 ;R ;i)
\cdots
\mathbb{TO}^{(2)}(m_{s},  N_{t}-1; n; R;i) 
\Biggr \} \; .
\eea
The average with respect to $m_{s}$ and $i$ provides
$\left [ P^{*} P\mathcal{O}(x,y; R) \right ]_{i_{c}}$.
For a schematic understanding, see Fig.~\ref{fig:LW-poly-ope}.
We repeat the same procedure as for the PLCF 
to get the final 
expectation value $\langle P^{*} P \mathcal{O} (x,y;R ) \rangle$.

\par
As a local operator, we use the electric field 
operator (parity odd) as
\be
\mathcal{O}_{E}(n)= i \bar{\theta}_{\mu\nu}(n)
= i ( \theta_{\mu\nu}(n)  - 2 \pi  n_{\mu\nu}(n)) \; ,
\label{eq:ope-ele}
\ee
where $\theta_{\mu\nu}(n)  \in [-4 \pi, 4\pi]$ 
and $n_{\mu\nu}(n) \in [0,\pm 1,\pm 2]$
is the modulo $2\pi$ of 
$\theta_{\mu\nu}(n)$, which corresponds 
to the magnetic  Dirac string.
Hence, one has $\bar{\theta}_{\mu\nu}(n) \in [-\pi, \pi ]$.
Moreover, we use the monopole current operator (parity odd) 
to detect the circulating monopole supercurrent as
\be
\mathcal{O}_{k}(\tilde{n})= 2 \pi i k_{\mu}(\tilde{n}) \; .
\label{eq:ope-mono}
\ee
Here $k_{\mu}$ is defined as the boundary of
the magnetic Dirac string~\cite{DeGrand:1980eq} as
\be
k_{\mu}(\tilde{n}) 
=
 \frac{1}{2} \varepsilon_{\mu\nu\alpha\beta}
\partial_{\nu} n_{\alpha\beta} (n+\hat{\mu}) 
\qquad  
\in [0,\pm 1, \pm 2] \; .
\ee
$\tilde{n}$ denotes the dual site
$n+ (\hat{1} +  \hat{2} + \hat{3} + \hat{4} )/2$.

\par
We have chosen these local operators
because of the possibility
to relate the U(1) flux tube and the classical 
flux-tube solution of the dual Ginzburg-Landau (DGL) theory.
In the DGL theory, the circulating 
monopole supercurrent induces the solenoidal electric field through
the dual Amp\`ere law, which plays a role in cancelling the 
Coulombic field induced by static charges at a distant place.
In this sense the measurement of the monopole current profile
is useful to judge whether the total electric flux is 
indeed squeezed or not.
In Appendix~\ref{sec:prof-decompose}, we show a
numerical evidence of such a cancellation mechanism 
of the electric flux inside the U(1) flux tube.
We may call this the composite structure
of the U(1) flux tube.

\par
Note that the definitions of the electric field 
and monopole current operators are not unique.
For instance, Cheluvaraja et al.~\cite{Cheluvaraja:2002yj}, 
have proposed an alternative
definition to satisfy the the Maxwell equations 
at finite lattice spacing.
It would also be interesting to study 
how the measured flux-tube profiles change with their operators.

\section{Numerical results : 
Static potential and force from the PLCF}
\label{sec:sec3}

In this section, we first present the simulation 
details associated with the LW algorithm
for the measurements of the 
potential and force from the PLCF, and then,
we show the corresponding numerical results.
Some analyses are also performed for the potential
and force; the potential is fitted with several ans\"atze.
The behavior of the force is
compared with the function derived from 
Eq.~\eqref{eqn:asympt-pot}.

\subsection{Simulation details}

We use a $16^{4}$ lattice. 
The $\beta$ values, the number of internal updates 
$N_{\rm iupd}$,
the number of configurations $N_{\rm conf}$, 
details of one Monte Carlo update (HB/OR),
and the range of measured distance between 
static charges $r/a$, are summarized in 
Table~\ref{tbl:measure1}.
Although we have not checked the finite volume effect,
as we will see later, 
our lattice volume itself is reasonably large even 
near the phase transition point:
$(\sim 3.5 r_{0})^{4}$  at $\beta = 1.01$. 
In addition to this, we restrict ourselves to measure the potential
up to $r/a = 6$.

\begin{table}[hbt]
    \centering
    \caption{Parameter setting}
    \begin{tabular}{|c|c|c|c|c|}
        \hline
        $\beta$ & $N_{\rm iupd}$ & $N_{\rm conf}$  & HB/OR
	&  $r/a$ range\\
        \hline
        0.98\F & 10000 & 1050 & 1 / 3 & 1--6\\
        \hline
        0.99\F & 8000\F & 1250 & 1 / 3  & 1--6 \\
        \hline
        1.00\F & 5000\F & 2000 & 1 / 3 & 1--6\\
        \hline
        1.005 & 3000\F & 2400 & 1 / 5  & 1--6 \\
        \hline
        1.01\F  &1000\F  & 3200 & 1 / 5  & 1--6 \\
        \hline
    \end{tabular}
    \label{tbl:measure1}
\end{table}

\par
\begin{figure}[t]
\centering
\includegraphics[height=9cm]{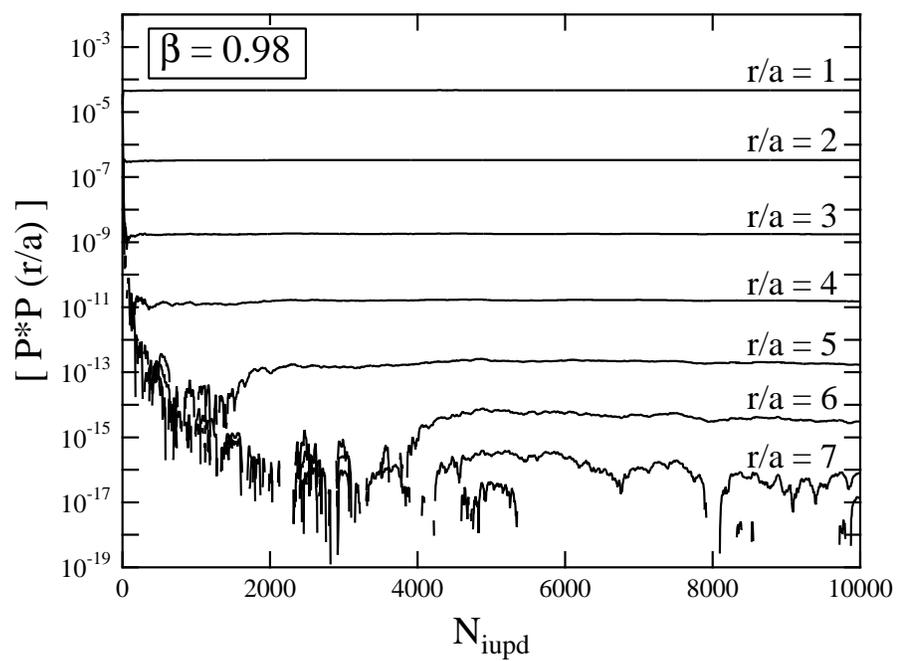}
\caption{Typical behavior of $[P^{*}P]$ for various $r$ 
at  $\beta$ = 0.98 as a function of $N_{\rm iupd}$.
When $N_{\rm iupd}$ is not sufficient,
$[P^{*}P]$ often takes negative values during the internal update,
typically for large $r$, where lines are broken.}
\label{fig:polyakov-sample}
\end{figure}

In Fig.~\ref{fig:polyakov-sample}, we show an example
how we have optimized $N_{\rm iupd}$ depending on $\beta$. 
This plot shows a typical behavior of a PLCF at $\beta=0.98$
for one configuration $[P^{*}P(r/a)]$ as a function of $N_{\rm iupd}$.
This figure tells us that, for instance, if we are interested 
in up to $r/a=4$, $N_{\rm iupd}=1000$ would be enough.
However, if $r/a=6$ is of interest, 
we may need to take $N_{\rm iupd} > 8000$.

\subsection{Static potential and force from the PLCF}

\par
In Table~\ref{tbl:alldata},
we summarize the expectation values of
the PLCF, potential and force for all $\beta$
values measured.
Note that it is possible to identify 
signals of the PLCF even when
$\langle P^{*}P \rangle = 10^{-3} \sim 10^{-16}$ with 
the $1 \sigma$ error varying from 0.4 to 8~\%.

\par
One may think that the investigation of the second 
derivative of the potential 
as in Refs.~\cite{Majumdar:2002mr,Luscher:2002qv}
for the direct identification 
of the coefficient of $1/r$ potential 
is also interesting.
However we have not succeeded to get reliable data for this.
The  result was strongly dependent on the definition of the
lattice second derivative.
This may be due to the reason that 
in U(1) LGT the rotational invariance is not well-recovered
compared to non-Abelian gauge theories even near 
the critical coupling.

\begin{table}[hbt]
    \centering
    \caption{The PLCF, static potential 
    and force.}
     \begin{tabular}{|c||c|c|c|c|c|}
       \hline
$\beta$  &   $r/a$ & $\langle P^{*}P(r/a) \rangle$ &
$aV(r/a)$ & $\bar{r}/a$ & $a^{2}F(\bar{r}/a)$  \\
       \hline
 0.98\F
 &  1 & 1.799(3)\F\F $\times \; 10^{-4\F}$ &  0.5389(1)\F & 1.5 & 
 0.4080(3)\F  \\
 &  2 & 2.627(14)\F $\times \; 10^{-7\F}$  & 0.9470(3)\F & 2.5 &  
 0.3439(5)\F  \\
 &  3 & 1.072(14)\F  $\times \; 10^{-9\F}$ & 1.2909(8)\F & 3.5 & 
 0.3150(6)\F  \\
 &  4 & 6.96(15)\F\F $\times \; 10^{-12}$  & 1.6058(14) & 4.5 &  0.3024(12)  \\
 &  5 & 5.54(17)\F\F  $\times \; 10^{-14}$ & 1.9083(19) & 5.5 &  0.3003(52)  \\
 &  6 & 4.77(35)\F\F $\times \; 10^{-16}$ & 2.2086(59) &  &   \\
       \hline
0.99\F
    &  1 & 3.186(6)\F\F $\times \;  10^{-4\F}$  &  0.5032(1)\F & 1.5 &  
    0.3562(3)\F  \\
    &  2 & 1.067(7)\F\F $\times \; 10^{-6\F}$ & 0.8594(4)\F & 2.5 &   
    0.2872(4)\F  \\
    &  3 & 1.080(14)\F $\times \; 10^{-8\F}$  & 1.1466(8)\F & 3.5 & 
    0.2585(7)\F  \\
    &  4 & 1.733(37)\F $\times \;  10^{-10}$ & 1.4051(14) & 4.5 & 
    0.2474(12)  \\
    &  5 & 3.35(13)\F\F  $\times \; 10^{-12}$ & 1.6525(24) & 5.5 &
    0.2444(40)  \\
    &  6 & 7.01(62)\F\F $\times \;  10^{-14}$ & 1.8970(59) &  &   \\
       \hline
  1.00\F
     &  1 & 6.507(11)\F $\times \; 10^{-4\F}$  &  0.4586(1)\F & 1.5 & 
     0.2920(2)\F  \\
     &  2 & 6.086(27)\F $\times \; 10^{-6\F}$ &   0.7506(3)\F & 2.5 &   
     0.2193(5)\F  \\
     &  3 & 1.825(18)\F  $\times \; 10^{-7\F}$& 0.9699(6)\F & 3.5 &
     0.1907(14)  \\
     &  4 & 8.66(16) $\times \; 10^{-9\F}$ & 1.1606(12) & 4.5 & 
     0.1797(27)  \\
     &  5 & 4.94(22)\F\F  $\times \; 10^{-10}$ & 1.3404(20) & 5.5 & 
     0.1791(35)  \\
     &  6 & 3.00(23)\F\F $\times \; 10^{-11}$ & 1.5195(45) &  &   \\
       \hline
   1.005
       & 1 & 1.045(2)\F\F $\times  \; 10^{-3\F}$  &  0.4290(1) & 1.5 & 
       0.2504(2)\F  \\
     &  2 & 1.902(10)\F $\times  \; 10^{-5\F}$ &   0.6794(3) & 2.5 &   
     0.1770(4)\F  \\
     &  3 & 1.121(10)\F  $\times  \; 10^{-6\F}$& 0.8565(7) & 3.5 & 
     0.1498(6)\F  \\
     &  4 & 1.027(21)\F $\times  \; 10^{-7\F}$ & 1.0063(13) & 4.5 &  
     0.1389(12)  \\
     &  5 & 1.136(42)\F  $\times  \; 10^{-8\F}$ & 1.1452(22) & 5.5 &
     0.1350(35)  \\
     &  6 & 1.42(11)\F\F $\times  \; 10^{-9\F}$ & 1.2802(52) &  &   \\
   \hline
1.01\F  
   &    1 & 2.152(8)\F\F $\times  \; 10^{-3\F}$  &  0.3839(2)\F & 1.5 & 
   0.1882(3)\F  \\
   &    2 & 1.061(10)\F $\times  \; 10^{-4\F}$ &   0.5720(4)\F & 2.5 &   
   0.1150(3)\F  \\
   &    3 & 1.687(24)\F  $\times \;  10^{-5\F}$&  0.6871(7)\F & 3.5 &
   0.0891(5)\F  \\
   &    4 & 4.076(79)\F $\times  \; 10^{-6\F}$ & 0.7762(12) & 4.5 & 
   0.0784(9)\F  \\
   &    5 & 1.180(37)\F  $\times  \; 10^{-6\F}$ & 0.8546(15) & 5.5 & 
   0.0755(27)  \\
   &    6 & 3.79(22)\F\F $\times \;  10^{-7\F}$ &  0.9301(45) &  &   \\
       \hline
   \end{tabular}
     \label{tbl:alldata}     
\end{table}

\clearpage
\subsection{Analysis}

\par
In this subsection, in order to see the presence/absence
of the universal $\gamma/r$ correction and
of more higher order corrections to the static potential,
we fit the static potential assuming the 
several explicit forms which are close to 
Eq.~\eqref{eqn:asympt-pot}:
\bea
&&
V_{1}(r) = \sigma r + \mu + \frac{C}{r} \; ,\\
\label{eq:pot1}
&&
V_{2}(r)= \sigma r + \mu + \frac{\gamma}{r}\; ,\\
\label{eq:pot2}
&&
V_{3}(r)= \sigma r + \mu 
+ \frac{\gamma}{r}\left ( 1 + \frac{b}{r}\right )\; ,
\label{eq:pot3}
\eea
where $\gamma = -\pi /12 \sim 0.262$.
The form of $V_{3}$ is motivated 
by Ref.~\cite{Luscher:2002qv}.

\par
Before carrying out the fitting, the PLCF has been 
averaged in bins over intervals between 50 and 160 time units 
(representing 5000 and 16000 iterations) depending on 
$\beta$ values to reduce the autocorrelation.
The potential has been computed from the nested PLCF.
For each fitting function, the means of the fitting parameters 
have been determined from the minimum of the $\chi^{2}$ which 
is defined with the covariance matrix so as to take into account the
correlation among different $r$'s.
The errors of the fitting parameters have  been estimated 
from the distribution of the jackknife samples of the fitting
parameters.
The fit range has been fixed so that the mean is consistent with
that evaluated by using only the diagonal part of the covariance 
matrix.

\par
The fitting results are summarized in 
tables~\ref{tbl:potfit1}, \ref{tbl:potfit2} and~\ref{tbl:potfit3}.
In Fig.~\ref{fig:potall_fit} we have plotted the potential and
the various fitting curves.
As seen from this figure all curves are on the data and practically
indistinguishable  from each other apart from the point $R=1$,
which lies outside of our fitting range.
In all cases the orders of the $\chi^{2}/N_{\rm DF}$ 
are one or smaller than one.
The coefficient $C$ obtained from  $V_1(r)$ is slightly different from
the theoretical value of $\gamma$, but not by much. 
However this maybe due to the existence of the Coulombic
$1/r$ potential in the current fitting region. 
An indication of this is seen in the fit by the potential $V_2(r)$
where we have had to drop the point $R=2$. 
$V_3(r)$ gives a better fit,  but of course it also contains 
an extra parameter.

\par
At this stage, one may wonder about the relevance of $1/r$ term 
at long distances, since all investigated types of the function
fit the potential very well.
To test this we have fitted the potential with the simple form 
$V_{4}(r)=\sigma r +\mu$ with $r=[3,6]$ and 
compared with the result of $V_{2}$.
It turned out that the fit was distinctly worse; 
the $\chi^{2}$ became 10 $\sim$ 20 times
larger than that of $V_{2}$ and the means were not consistent
with that determined from the use of the diagonal part of
the covariance matrix.

\par
\begin{figure}[!t]
    \centering
     \includegraphics[height=10cm]{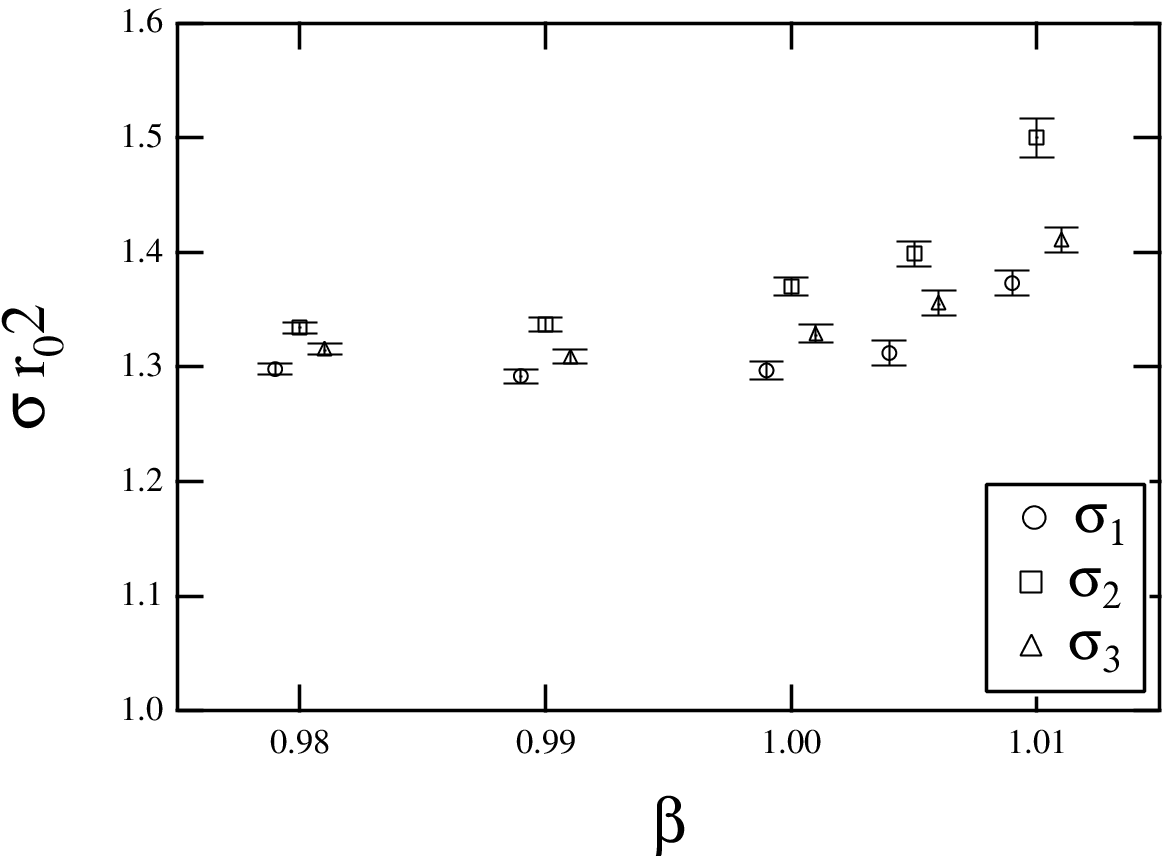}
    \caption{String tensions as a function of $\beta$.
    $\sigma$ corresponding to $V_{i}$ are denoted by $\sigma_{i}$.
    The bottom axis for $\sigma_{1}$ and $\sigma_{3}$
    are slightly shifted to distinguish each other.}
     \label{fig:st_beta}
      \includegraphics[height=10cm]{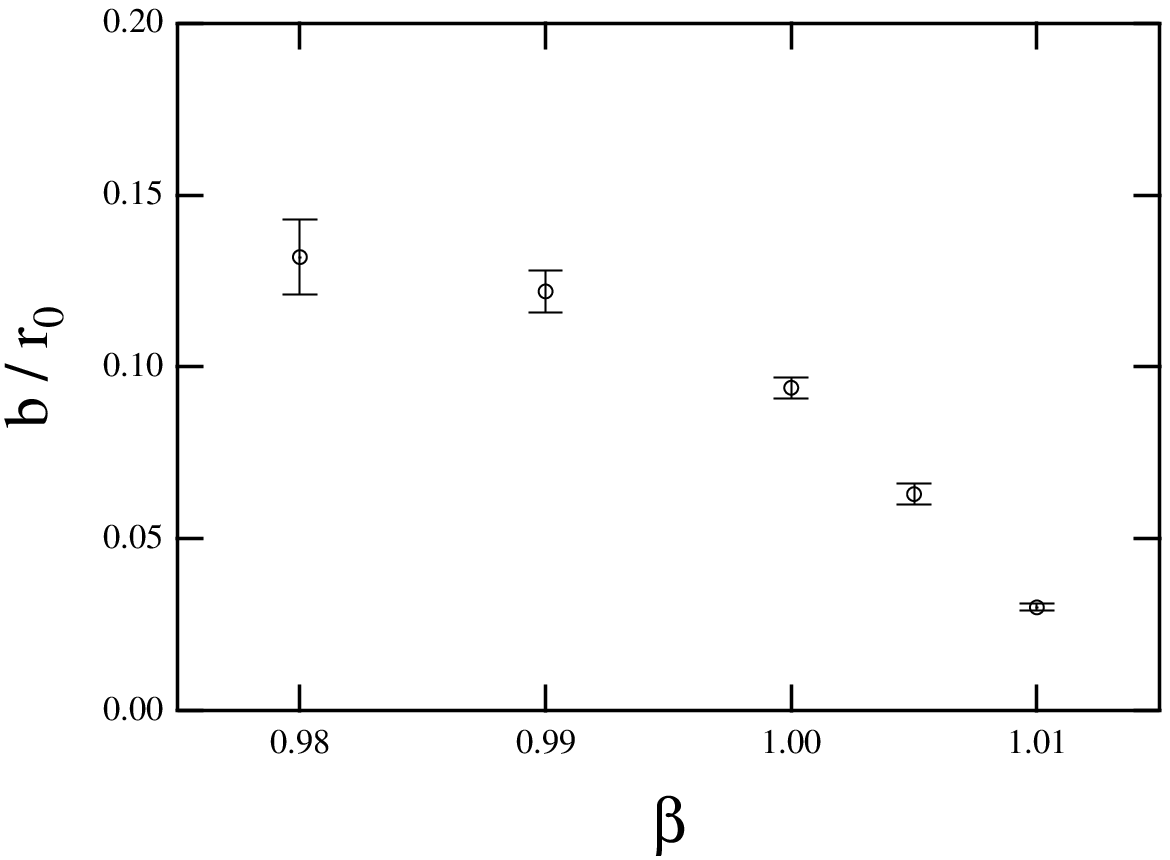}
      \caption{$b/r_{0}$ as a function of 
      $\beta$ (see, $V_{3}(r)$ in Eq.~\eqref{eq:pot3}).}
     \label{fig:b3_beta}
\end{figure}
\begin{figure}[hbt]
     \centering
      \includegraphics[height=10cm]{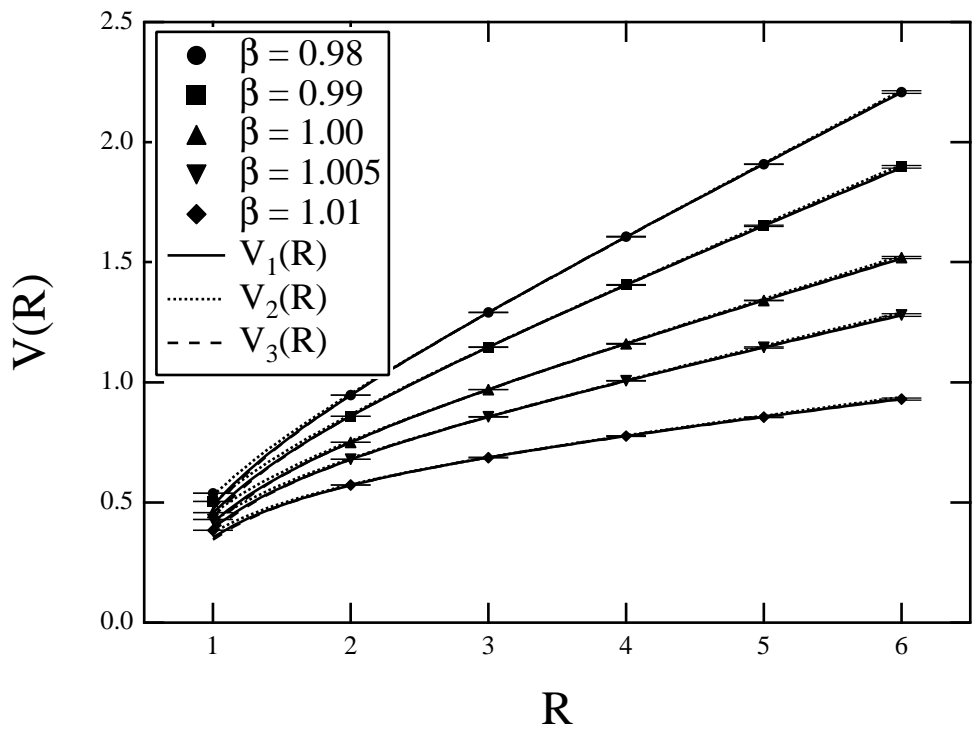}
     \caption{Potentials for various $\beta$ vs. the fitting curves}
      \label{fig:potall_fit}
 \end{figure}

To compare the fit results among different $\beta$
values, we introduce a scale $r_{0}$ 
based on Sommer's relation 
$F(r_{0}) r_{0}^{2} = 1.65$~\cite{Sommer:1994ce}.
We have found 
$r_{0}/a=2.13$, 2.37, 2.83(1), 3.28(1), 4.60(3)
for $\beta=0.98$, 0.99, 1.00, 1.005 and 1.01, respectively.
For first two $\beta$ values there were no error at this order.
Using $r_{0}$, we plot $\sigma r_{0}^{2}$ 
from $V_{1}$, $V_{2}$ and $V_{3}$
as a function of $\beta$ in Fig.~\ref{fig:st_beta}.
The bottom axis corresponding to
$V_{1}$ and $V_{3}$ are slightly shifted in the plot
to distinguish them from each other.
We see slight differences among
the string tensions for different ans\"atze of the potential.
The string tensions at $\beta=$0.98 and 0.99 show a
good scaling behavior with respect to $r_{0}$.
For $\beta > 1.00$, $ \sigma_{i} r_{0}^{2}$  start 
to grow, which suggests that the 
approach to the phase transition point
of the string tensions and $r_{0}$ are different.
In Fig.~\ref{fig:b3_beta}, we plot $b/r_{0}$ against $\beta$,
where $b$ is a coefficient of the $1/r^{2}$ potential
in Eq.~\eqref{eq:pot3}.
It is interesting to find that at $\beta=$0.98 and 0.99, 
$b/r_{0}$ seems to be saturated.
As $\beta$ increases, however, it falls down to zero. 
Since for large $\beta$ we can look at only
the short range of the potential, this result may  
suggest that $b$ is irrelevant for such a range.

\par
\begin{figure}[t]
    \centering
     \includegraphics[height=10cm]{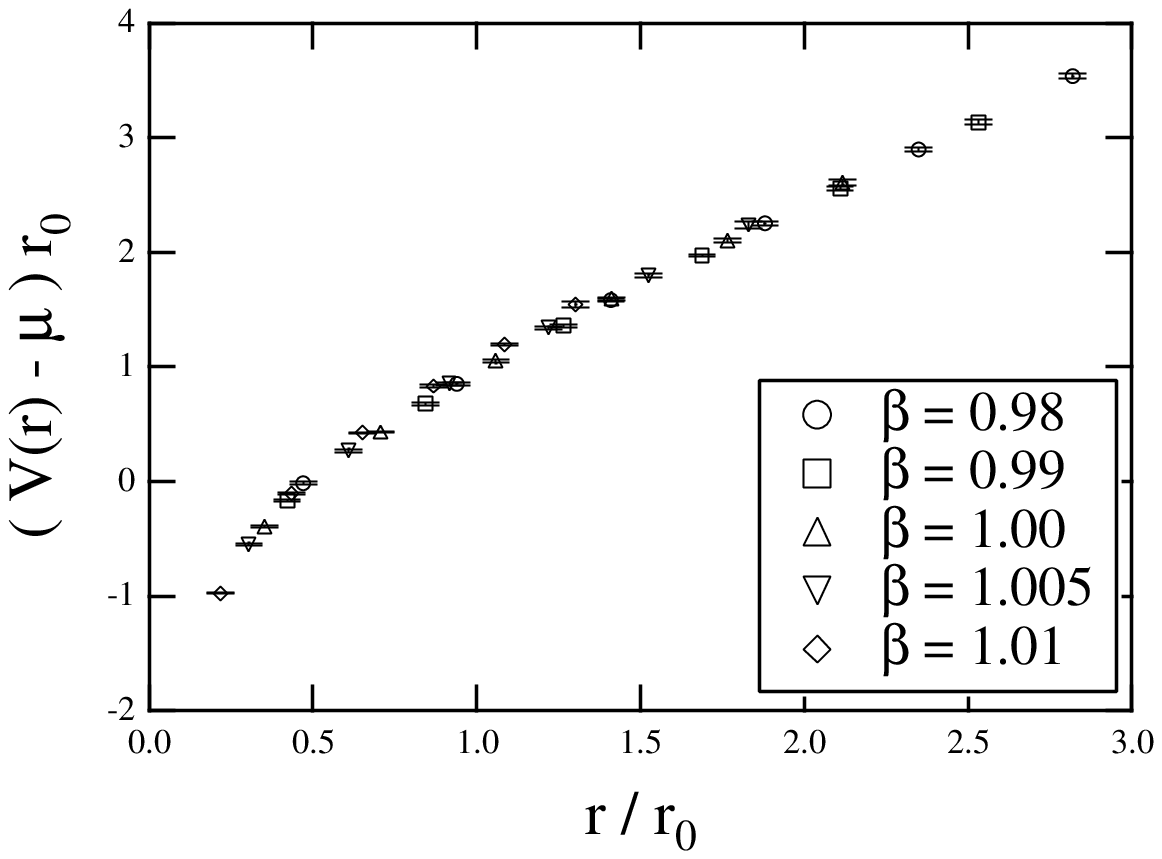}
    \caption{Static potential as a function of $r/r_{0}$. 
    Constant $\mu$ is determined by $V_{1}$ fit.}
     \label{fig:potential_r0}
      \includegraphics[height=10cm]{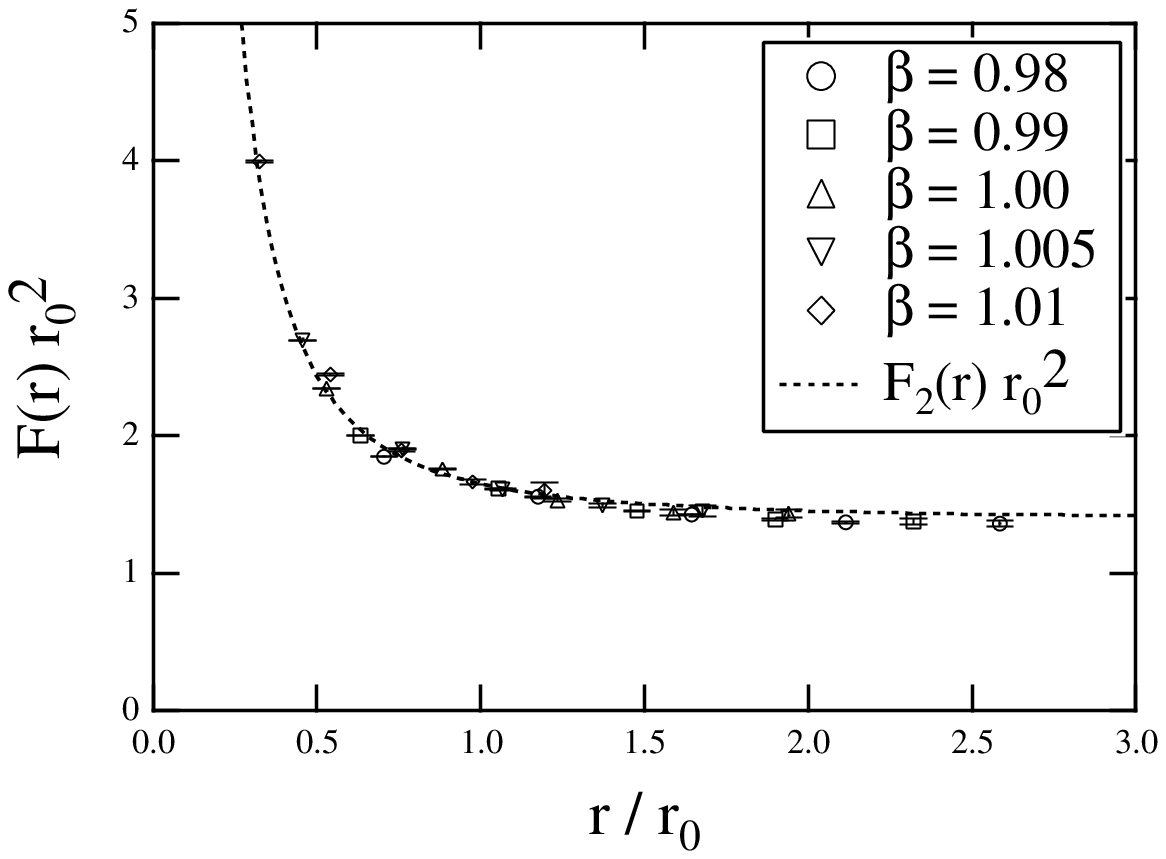}
    \caption{Force as a function of  $r/r_{0}$. 
    The dashed line corresponds to 
    $F_{2}(r)=dV_{2}(r)/dr=\sigma- \gamma/r^{2}$
    with $\sigma=(1.65-\pi/12)/r_{0}^{2}$.}
     \label{fig:force_r0}
\end{figure}

We then show the potential for all $\beta$
as a function of $r/r_{0}$ in Figs.~\ref{fig:potential_r0}.
To subtract the constant $\mu$ in the potential plot, 
we have used the value obtained by the $V_{1}$ fit.
We find that the potential beautifully falls onto 
one curve except for the data from $\beta=1.01$.
The reason for this exception can be understood from 
Fig.~\ref{fig:st_beta}, which shows that the string tension 
$\sigma r_{0}^{2}$ grows as the $\beta$ approaches to 
the phase transition point.
We remark that the result was insensitive to the 
choice of the potential form used in the fit
(corresponding figure from the $V_{2}$ fit is
found in Ref.~\cite{Koma:2003gv}).

\par
In Fig.~\ref{fig:force_r0}, we plot the
force for all $\beta$ as a function of $r/r_{0}$ and the
expected function from the potential $V_{2}$:
$F_{2}(r) \equiv dV_{2}(r)/dr=\sigma -\gamma/r^{2}$.
It should be noted that this function contains no
fitting parameter, since Sommer's relation 
gives a fixed value for the string tension
$\sigma r_{0}^{2}  =1.65 -\pi/12 \sim  1.39$.
We find that the general behavior of the force data seems to be
described by this function.
Although there are slight differences between the curve and the 
data at long distance,
we consider that this result supports the universality of $\gamma/r$ 
correction to the static potential.
To make a more precise statement, however, one has to control
various systematic effects which enter in this analysis,
for instance, 
the contribution of $1/r^2$ force which originates from the 
Coulombic electric field  (partially discussed in 
Appendix~\ref{sec:prof-decompose}), and
the possibility of higher order corrections to the static 
potential, which are not universal and vary from theory to 
theory.
Surprisingly, another feature in common with non-Abelian
gauge theories~\cite{Weisz:1995} is that this 
function also fits the data down to 
relatively short distance to $r/r_{0} \sim 0.3$.
For this, there is as yet  no explanation.

\begin{table}[hbt]
    \centering
     \caption{Potential fit by 
    $V_{1}(r) = \sigma r + \mu + \frac{C}{r}$.}
\begin{tabular}{|c||c|c|c|c|}
    \hline
$\beta$ & $\sigma a^{2}$  & $\mu a$ & $C$ &  fit range ($r/a$) \\
    \hline
0.98\F & 0.286(1) & 0.546(5) & $-$0.344(6) & 2$-$6   \\
\hline
0.99\F & 0.230(1) & 0.573(4) & $-$0.346(4)& 2$-$6  \\
\hline
1.00\F & 0.162(1)  & 0.598(3) & $-$0.342(3)& 2$-$6   \\    
\hline
1.005 & 0.122(1) & 0.597(3) & $-$0.326(3)& 2$-$6\\
    \hline
1.01\F & 0.0649(5) &0.595(1) & $-$0.305(1) & 2$-$6 \\ 
     \hline
\end{tabular}
\label{tbl:potfit1}
\end{table}

\begin{table}[hbt]
    \centering
  \caption{Potential fit by 
    $V_{2}(r)= \sigma r + \mu + \frac{\gamma}{r}$ with 
    $\gamma=-\pi/12$.}
\begin{tabular}{|c||c|c|c|}
    \hline
 $\beta$ & $\sigma a^{2}$  & $\mu a$ &  fit range ($r/a$) \\
    \hline
 0.98\F &   0.294(1)  & 0.497(2) &  3$-$6 \\
 \hline
  0.99\F &  0.238(1)  & 0.521(2)&  3$-$6   \\
    \hline   
1.00\F &    0.171(1)  &  0.545(1) &  3$-$6  \\
    \hline   
1.005 &   0.130(1) & 0.554(1) &  3$-$6  \\
    \hline   
1.01\F &    0.0709(8) & 0.565(1) &  3$-$6  \\
    \hline   
\end{tabular}
\label{tbl:potfit2}     
\end{table}

\begin{table}[hbt]
    \centering
    \caption{Potential fit by 
    $V_{3}(r)= \sigma r + \mu 
    + \frac{\gamma}{r}\left ( 1 + \frac{b}{r}\right )$ with 
    $\gamma=-\pi/12$.}
\begin{tabular}{|c||c|c|c|c|}
   \hline
 $\beta$ & $\sigma a^{2}$  & $\mu a$ & $b/a$ 
 &  fit range ($r/a$) \\
    \hline
0.98\F &  0.290(1) & 0.517(4) & 0.281(23)  & 2$-$6   \\
     \hline
0.99\F & 0.233(1) & 0.543(3)  & 0.289(15)  & 2$-$6   \\
     \hline
1.00\F &   0.166(1) & 0.567(2)  &0.267(9)  & 2$-$6    \\
        \hline
1.005 & 0.126(1) & 0.572(2) & 0.208(10)  & 2$-$6   \\
       \hline
1.01\F  &0.0667(5) & 0.578(1) & 0.138(5) & 2$-$6  \\
     \hline
\end{tabular}
  \label{tbl:potfit3} 
\end{table}

\clearpage
\section{Numerical results : Flux-tube profile}
\label{sec:sec4}

In this section, we show numerical results
on the flux-tube profile.
We then investigate how the width of the flux tube
behaves as a function of $r$ based on the 
DGL analysis.

\subsection{Simulation details}

The $\beta$ values, the lattice volume, and
details of one  Monte Carlo update (HB/OR)
are the same as the measurement of the PLCF.
Since in this case we do not compute derivatives 
with respect to $r$, we have changed $N_{\rm iupd}$ 
depending on $r$ in order 
to achieve a reasonable performance.
The number of $N_{\rm iupd}$ is
summarized in Table~\ref{tab:profile-conf}.
We have measured the profile on 
the mid-plane between charges as described 
in subsection~\ref{subsec:procedure2}.
In order to compare the profile among different $\beta$'s
and $r$'s easily, we take the cylindrical average
of the 2D profile; we define the radius 
$\rho=\sqrt{x^{2}+y^{2}}$ and the azimuthal angle
around $z$ axis as $\varphi = \tan^{-1}(y/x)$.

\begin{table}[hbt]
\caption{The number of $N_{\rm iupd}$ for the measurement
of the flux-tube profile.
We gave up the profile measurement for $r/a=6$ at $\beta=0.98$
because of a practical reason.}
\centering
\begin{tabular}{|c||c|c||c|c||c|c||c|c||}
    \hline
    $\beta$  &  $r/a$  & $N_{\rm iupd}$  & $r/a$  & $N_{\rm iupd}$ 
    & $r/a$  & $N_{\rm iupd}$   & $r/a$  & $N_{\rm iupd}$   \\
    \hline
0.98\F   & 3 & 200 & 4 & 1000 & 5 & 8000 & 6 & $---$ \\
    \hline
0.99\F   & 3 & 200 & 4 & 1000 & 5 & 5000 & 6 & 8000 \\
   \hline
1.00\F   & 3 & 200 & 4 & 1000 & 5 & 3000 & 6 & 5000 \\
    \hline  
1.005 & 3 & 200 & 4 & 1000 & 5 & 2000 & 6 & 3000 \\
    \hline
1.01\F   & 3 & 200 & 4 & 1000 & 5 & 1000 & 6 & 1000 \\
    \hline  
\end{tabular}
\label{tab:profile-conf}
\end{table}

\subsection{Flux-tube profile}

\begin{figure}[hbt]
\centering
\includegraphics[width=13cm]{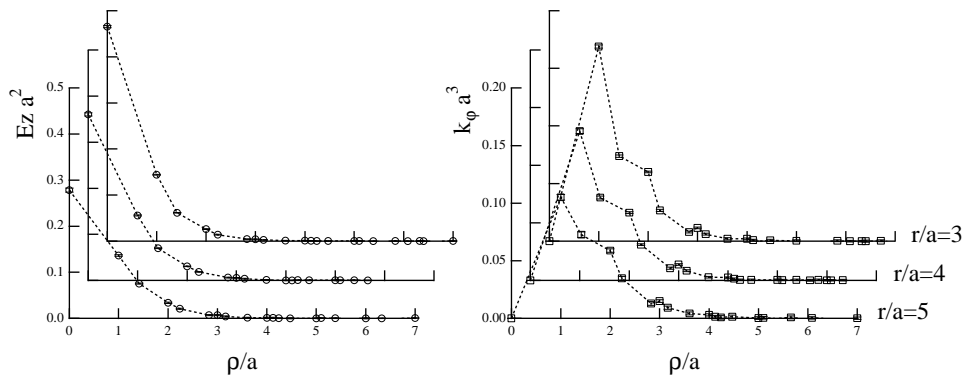}
\caption{Profiles of the electric field (left) and of
the monopole current (right) for $r/a=$3, 4, 5
at $\beta=0.98$.}
\label{fig:b098-string}
\end{figure}

\begin{figure}[hbt]
\centering
\includegraphics[width=13cm]{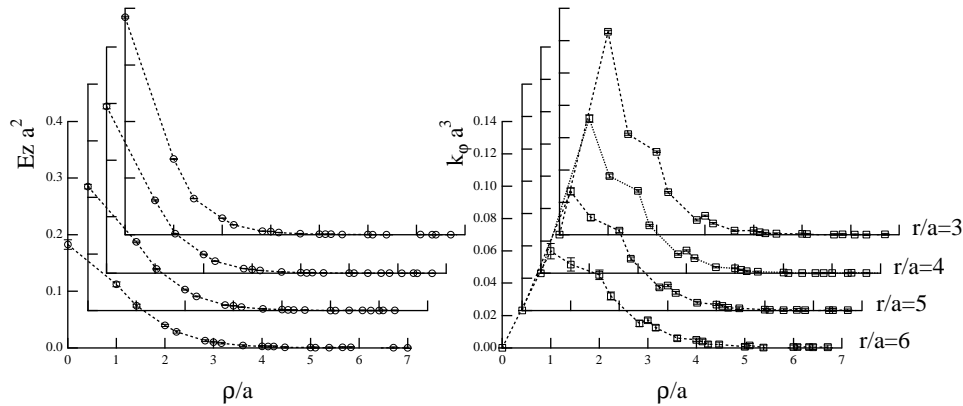}
\caption{The same plot as in Fig.~\ref{fig:b098-string}
at $\beta=0.99$.}
\label{fig:b099-string}
\end{figure}

\begin{figure}[hbt]
\centering
\includegraphics[width=13cm]{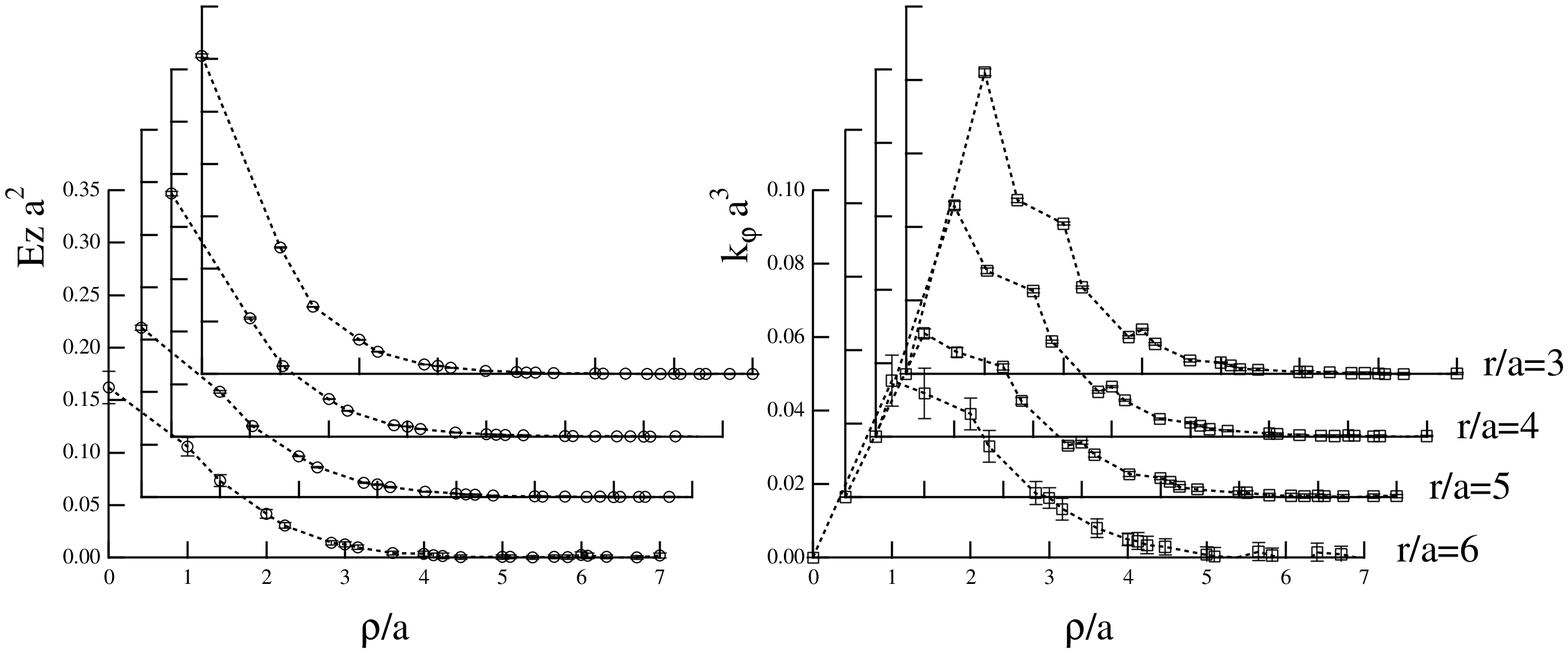}
\caption{The same plot as in Fig.~\ref{fig:b098-string}
at $\beta=1.00$.}
\label{fig:b100-string}
\end{figure}

\begin{figure}[t]
\centering
\includegraphics[width=13cm]{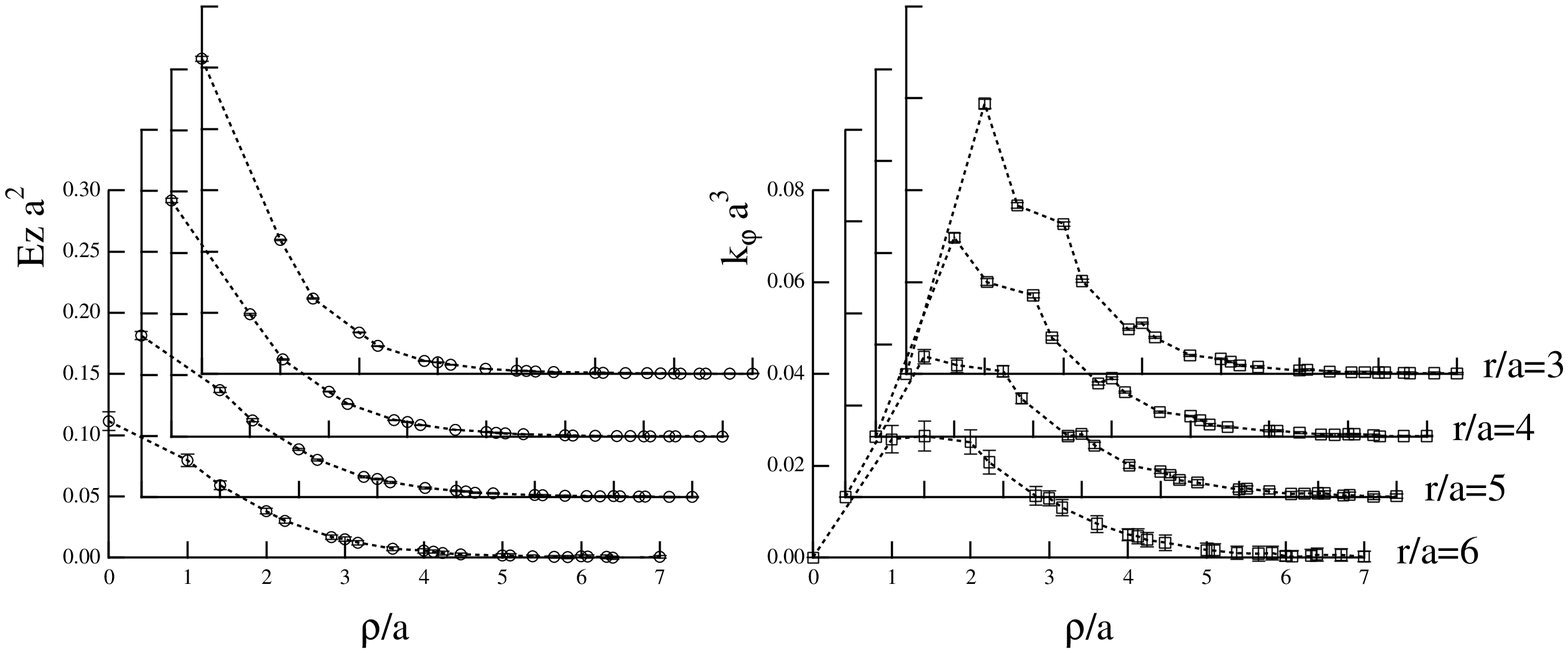}
\caption{The same plot as in Fig.~\ref{fig:b098-string}
at $\beta=1.005$.}
\label{fig:b1005-string}
\end{figure}

\begin{figure}[t]
\centering
\includegraphics[width=13cm]{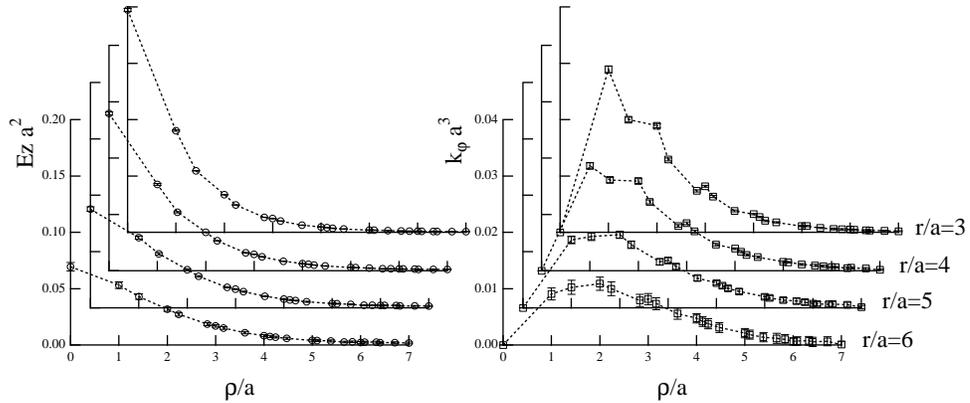}
\caption{The same plot as in Fig.~\ref{fig:b098-string}
at $\beta=1.01$.}
\label{fig:b101-string}
\end{figure}

We show the profiles of electric field and monopole current
in Figs.~\ref{fig:b098-string}~--~\ref{fig:b101-string}.
The number of configurations is $N_{c}=300$ for all data.
The investigated ranges are  
$r/r_{0}=0.469-2.35$ at $\beta=0.98$,
$r/r_{0}=0.422-2.53$ at $\beta=0.99$,
$r/r_{0}=0.353-2.12$ at $\beta=1.00$,
$r/r_{0}=0.305-1.83$ at $\beta=1.005$
and
$r/r_{0}=0.217-1.30$ at $\beta=1.01$.
At glance we find that all data are clean enough, which 
allow us to identify the profile.

\par
For all $\beta$ values we find a tendency that
as $r$ increases, the peak of the electric field at 
$\rho \sim 0$ decreases and the strong peak of the 
monopole current profile for a small $\rho$ disappears.
This feature can be understood as follows.
For a small $r$, the mid-plane is close to
the static charges so that the 
Coulombic electric field contribution is still large.
In order to cancel such a strong electric field as much as possible,
the monopole supercurrent must be high, which is indicated 
by the strong peak in 
Figs~\ref{fig:b098-string}~--~\ref{fig:b101-string}.
For larger $r$ the Coulombic field is weaker in the middle
and consequently the peak of the monopole current 
is also weaker. 
It is interesting that although the rotational 
invariance does not hold for the monopole current profile, 
especially for small $r$, it is effectively restored as $r$ increases.
 
\clearpage
\subsection{Analysis}

In this subsection, we investigate how the width of 
the flux-tube profile depends on $r$ based on the DGL analysis.
In particular, we pay attention to the data for $r/a=5$.
Corresponding flux-tube lengths are $r/r_{0}=1.09 \sim 2.35$.
The number of configuration used for this analysis
is $N_{\rm conf}=400$, 500, 700, 900 and 1200 
for $\beta=0.98$, 0.99, 1.00, 1.005 and 1.01, respectively.
We have increased $N_{\rm conf}$ for larger $\beta$ values 
to compensate the enhancement of statistical errors 
due to the smaller lattice spacings when the reference 
scale is introduced.

\par
\begin{figure}[!t]
\centering
\includegraphics[height=10cm]{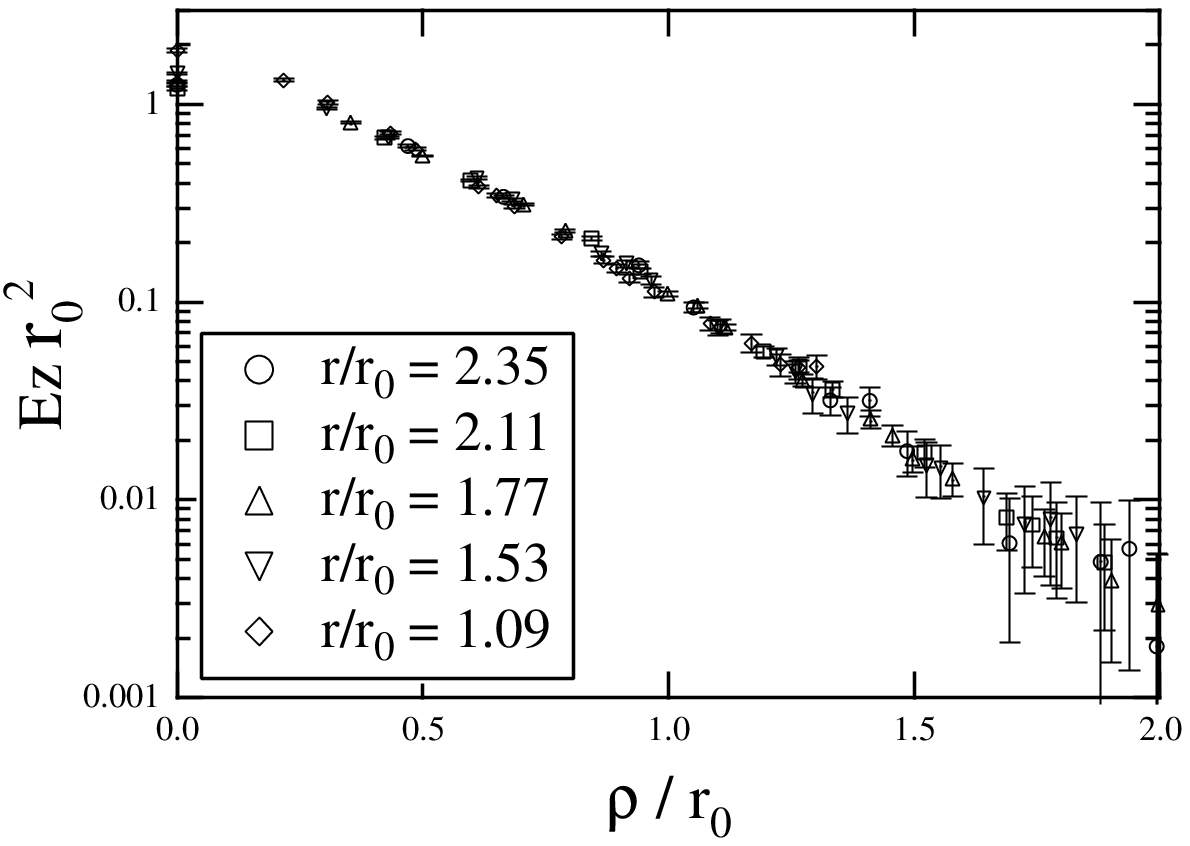}
\caption{Profiles of the electric field from different $r$.}
\label{fig:electric_scale}
\includegraphics[height=10cm]{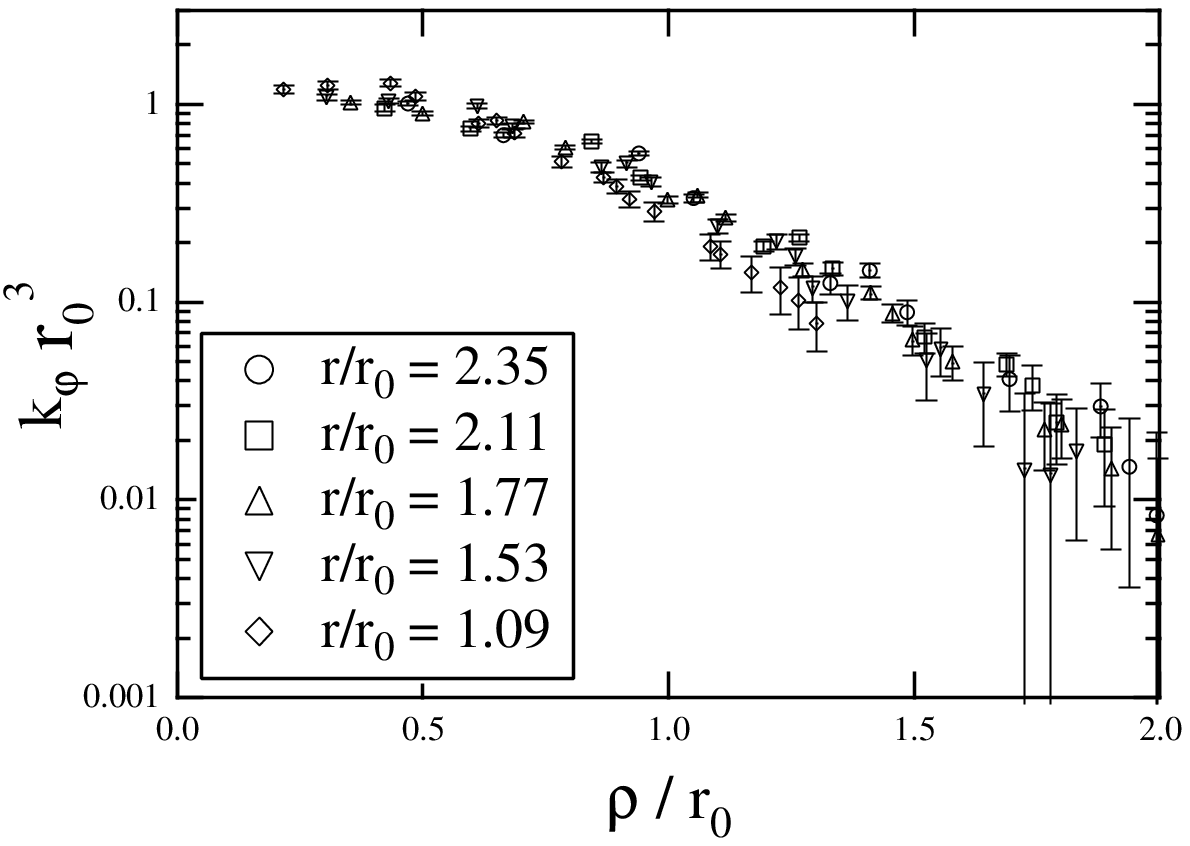}
\caption{Profiles of the monopole current from 
different $r$.}
\label{fig:monopole_scale}
\end{figure}

Before the analysis, let us put all profiles into one figure
by introducing the Sommer scale $r_{0}$.
In Figs.~\ref{fig:electric_scale} 
and~\ref{fig:monopole_scale}, we show the profiles
of the electric field and of the monopole current, 
respectively.
We find that the tail of the electric field profile 
$(\rho/r_{0} > 0.4)$ seems to fall into one curve,
indicating its $r$ independence.
Moreover, although the monopole current profile shows 
squeezed shape for small $r$, it
becomes wider as increasing $r$
and seems to converge again into one curve. 
Therefore, in the following DGL analysis 
we particularly pay attention to the tail behavior of 
the profiles.

\par
We briefly describe the DGL theory and the
classical flux-tube solution used for this analysis.
The DGL Lagrangian density is given by
\be
\mathcal{L}_{DGL}
= -\frac{1}{4}
(\partial_{\mu}B_{\nu}
-\partial_{\nu}B_{\mu} - e {}^{*\!}\Sigma_{\mu\nu})^{2}
\! + \!
| (\partial_{\mu} \!+ \!  ig B_{\mu})\chi |^{2}
\! -\! \lambda ( |\chi|^{2}\!  -\! v^{2})^{2} \; ,
\label{eqn:dgl}
\ee
where $B_{\mu}$ and $\chi=\phi e^{i \eta}$ $(\phi, \eta  \in \Re)$
are the dual gauge field and the monopole field.
$\Sigma_{\mu\nu}$ describes an external electric Dirac 
string sheet, which is bounded by the electric current
$\partial_{\mu}\Sigma_{\mu \nu}= j_{\nu}$.
This singularity is responsible for 
the location and length of the flux tube, which determines
the singular part of the dual gauge field~\cite{Koma:2000wn}.
The masses of the dual gauge boson and monopole 
can be expressed as $m_{B}=\sqrt{2}g v$ and 
$m_{\chi}=2 \sqrt{\lambda}v$, respectively.
The inverses of these masses are corresponding to the 
penetration depth and coherence length, which 
characterize the width of the flux tube.
Note that the electric coupling $e$ and the dual gauge coupling
$g$ satisfy the Dirac quantization condition $eg=2 \pi$.
We consider a translational invariant flux tube 
along the $z$ axis by parameterizing the system
with cylindrical coordinate $(\rho, \varphi, z)$.
As mentioned, due to the Dirac string $\Sigma_{\mu \nu}$ in
the dual field strength,
the dual gauge field consists of a regular and a singular 
parts as 
$B_{\mu}=B_{\mu}^{\rm reg}+B_{\mu}^{\rm sing}$.
For the given system, each part is reduced to
$\bvec{B}^{\rm reg}=\tilde{B}(\rho)/\rho \; \bvec{e}_{\varphi}$
and  $\bvec{B}^{\rm sing}= -1/(g\rho) \; \bvec{e}_{\varphi}$.
The field equations for $\tilde{B}(\rho)$ and $\phi(\rho)$ 
are then given by
\bea
&&
\frac{d}{d\rho} 
\left ( 
\frac{1}{\rho} \frac{d\tilde{B}}{d \rho}
\right )
-2g \phi^{2} \left ( \frac{g \tilde{B}-1}{\rho}  \right )
=0\; ,\\*
&&
\frac{d^{2}\phi}{d\rho^{2}}+ \frac{1}{\rho}
\frac{d \phi}{d \rho}- 
\left (\frac{g\tilde{B}-1}{\rho}\right ) \phi
- 2 \lambda \phi (\phi^{2}-v^{2})=0 \; .
\eea
The second term of the first equation is identified as 
the azimuthal monopole current 
$\bvec{k}= k(\rho)\bvec{e}_{\varphi}$ with 
$k(\rho)=-2g \phi^{2} ( g \tilde{B}-1)/\rho$.
One can solve the field equations analytically at  
large radius~$\rho$ where~$\phi \sim v$.
In such region, the second equation provides the 
boundary condition of $\tilde{B}$  as $\tilde{B} \to 1/g $.
By writing $\hat{\rho}= m_{B} \rho$ and 
$\tilde{B}=1/g - \hat{\rho}K (\hat{\rho})$,
the first equation can be rewritten as
\bea
\frac{d^{2}K}{d \hat{\rho}^{2}}
+\frac{1}{\hat{\rho}}\frac{d K}{d \hat{\rho}}
-\left ( 1 +\frac{1}{\hat{\rho}^{2}} \right ) K = 0 \; .
\eea
The solution is the first-order modified Bessel function
$K = K_{1} (\hat{\rho})$.
For large $\rho$ it behaves as 
$K_{1}(\hat{\rho}) \sim \sqrt{\frac{\pi}{2 \hat{\rho}}}
e^{-\hat{\rho}}$.
Using this, one finds the solution
for the electric field and the monopole current as
\bea
&&
E_{z}( \hat{\rho}) =\frac{1}{\rho}
\frac{d \tilde{B}}{d \rho}
= m_{B}^{2}K_{0}(\hat{\rho}) \; ,\\
\label{eqn:fit-e}
&&
k(\hat{\rho}) = m_{B}^{3} K_{1} (\hat{\rho})  \; .
\label{eqn:fit-k}
\eea
We use these functions to find $m_{B}$.
However, we must keep in mind that this solution is
applicable only the region 
where the system is translational invariant and
the monopole field has a vacuum expectation value $\phi \sim v$. 

\par
We have employed a similar fitting procedure as used in 
the potential fit.
Before carrying out the fitting, the 
$\langle  P^{*}P \mathcal{O}  \rangle$,
where $\mathcal{O}$ denotes a local operator, and $\langle  P^{*}P\rangle$ 
have been averaged in bins over intervals between 20 and 40 time 
units (representing 2000 and 4000 iterations) depending on 
$\beta$ values to reduce the autocorrelation.
Then, the cylindrical profile composed only from the on-axis data
has been computed.
The mean of the dual gauge boson mass has
been determined from the minimum of $\chi^{2}$
defined with the diagonal part of the covariance matrix, 
since the correlation among different $\rho$'s
was not significant,
which may be due to the cylindrical averaging.
The error has been estimated from the distribution of the
jackknife samples of the fitting parameters.
The fit range has been fixed so that the mean and/or 
the order of $\chi^{2}$ is stable against the change of the 
fit range.

\par
\begin{figure}[hbt]
\centering
\includegraphics[height=10cm]{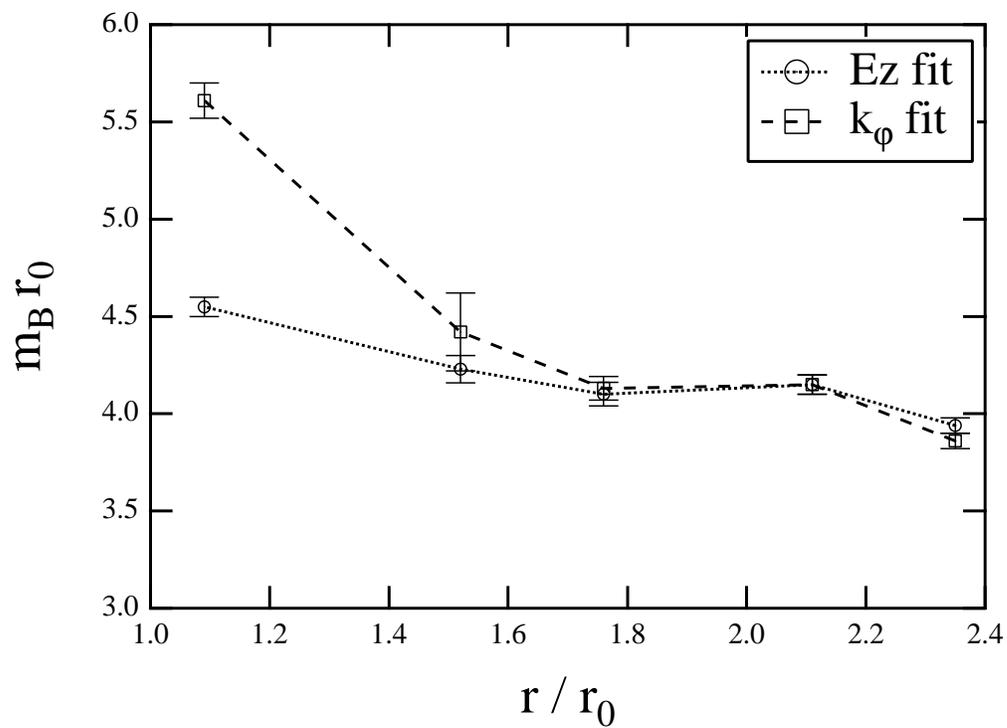}
\caption{Dual gauge boson mass
extracted from the fit as a function of $r$.}
\label{fig:mb_rdep}
\end{figure}

In Table~\ref{tbl:fit-mb},  we summarize the result.
The minimum radii which satisfy the above condition are found
to be $\rho_{\rm min}/a=3$ for $\beta=0.98$ and 0.99,
$\rho_{\rm min}/a=4$ for $\beta=1.00$ and 1.005,
and $\rho_{\rm min}/a = 5$ for $\beta=$1.01.
In Fig.~\ref{fig:mb_rdep}, we plot $m_{B}r_{0}$ as a 
function of $r/r_{0}$.
We find that while the mass extracted from the $k_{\varphi}$
fit for small $r$ is larger than that from $E_{z}$ fit, 
it approaches $E_{z}$'s result with increasing~$r$.
Basically, if the ansatz for the DGL flux-tube solution
(translational invariance along $z$ axis and $\phi \sim v$) is valid,
the masses extracted from $E_{z}$ and those from $k_{\varphi}$ 
fits should coincide with each other.
In this sense we should take only the result
for $r/r_{0} \geq 1.77$ seriously.
In fact, for short distances $k_{\varphi}$ cannot be 
translational invariant, because it is responsible for the
solenoidal electric field inside a flux tube, which 
cancels the Coulombic field at
large~$\rho$ region (see, Appendix~\ref{sec:prof-decompose}).
We find that the mass is stable around $m_{B}r_{0} \sim 4.0$.
If the width of the flux tube diverges according to the 
prediction of the effective bosonic string theory, 
the dual gauge boson mass, identified within this 
analysis, should go to zero (penetration depth goes to infinity).
However, we have observed an almost constant
behavior of the width of the 
flux tube in this range.
Whether a logarithmic growth of the width
is hidden in our data, especially for the data at $r/r_{0}=2.35$,
is however difficult to say.

\par
Finally, we would like to discuss further the
detailed fit to investigate all the three parameters 
in the DGL theory.
We have performed the full range profile fit 
(whole $\rho$ region but only the on-axis data) 
using the finite length flux-tube solution 
obtained numerically within the 3D lattice discretized DGL 
theory as in Ref.~\cite{Koma:2003hv}.
Here, both the electric field and the monopole
current profiles have been fitted simultaneously,
where the diagonal part of the covariance matrix has been
taken into account to define the $\chi^{2}$.
The procedure to estimate the error has been the same as above.
We have found, however,  a tendency that 
this method cannot reliably be applicable
for $\beta = 0.98 - 1.005$, where
the fine structure of the monopole current around the
peak is not clear due to the large lattice spacing.
In these cases, we could not identify the minimum 
of $\chi^{2}$ within the three parameter space,
especially along the axis of the monopole mass.
Although the dual gauge coupling has been almost a constant
$\beta_{g} \sim 0.06$, the dual gauge boson mass
and the monopole mass have been correlated to each other.
The problem is that many sets of mass parameters
have reproduced the profile well 
and due to this we could not find an unique set of parameters.
In fact, for instance, as seen in Fig.~3 of 
Ref.~\cite{Koma:2003gq},
the monopole mass is responsible for the 
shape of the monopole current profile in small $\rho$ region,
while the dual gauge boson mass that in large $\rho$ region.
Only the monopole current profile at $\beta=1.01$ 
(see, Fig.~\ref{fig:b101-string} or 
Fig.~\ref{fig:monopole_scale} at $r/r_{0}=1.09$)
has the peak at $\rho/a=2 > 1$.
In this case we could find a minimum of $\chi^{2}$ 
and the parameters were found to be
$\beta_{g}=0.061(2)$, $m_{B}a=0.59(2)$ 
and $m_{\chi}a=0.59(6)$.
Taking into account the fact that the lattice spacing is 
still large at $\beta=1.01$, we may say that 
these masses are consistent
with the glueball masses in the axial-vector and scalar
channels  given in Ref.~\cite{Majumdar:2003xm}.

\begin{table}[hbt]
\centering
\caption{The dual gauge boson mass extracted from the fit.}
\begin{tabular}{|c|c|c|}
    \hline
    $\beta$ & $m_{B}a$ & fit range ($\rho/a$) \\
    \hline
    0.98\F ($E_{z}$) &  1.85(2) &  3--6     \\
    0.98\F  ($k_{\varphi}$) &1.81(2)  & 3--6  \\
    \hline
     0.99\F ($E_{z}$) &  1.75(2) &  3--6  \\
    0.99\F  ($k_{\varphi}$) &1.75(2)  & 3--6 \\
    \hline
       1.00\F ($E_{z}$) &  1.45(2) &  4--6  \\
    1.00\F  ($k_{\varphi}$) &1.46(2)  & 4--6 \\
    \hline
      1.005 ($E_{z}$) &  1.29(2)   &  4--6   \\
    1.005  ($k_{\varphi}$) &1.35(6)  & 4--6 \\
    \hline
     1.01\F ($E_{z}$) &  0.990(11)   &  5--6   \\
    1.01\F  ($k_{\varphi}$) &1.22(2)  & 5--6   \\
    \hline
\end{tabular}
  \label{tbl:fit-mb} 
\end{table}

\clearpage
\section{Summary}
\label{sec:sec5}

We have successfully simulated the Wilson gauge action of 
4D compact U(1) lattice gauge theory on a  $16^{4}$ lattice 
at $\beta=$ 0.98, 0.99. 1.00, 1.005 and 1.01
(confinement phase) by using the multi-level algorithm.

\par
First, we have measured the static potential and 
force between two static charges from the Polyakov 
loop  correlation function (PLCF) up to the distance
$r_{\rm max}/r_{0}=2.82$.
It was possible to identify the PLCF which
take values from $10^{-3}$ 
to $10^{-16}$ within $10$ \% error.
We have analyzed the potential and force
by fitting with several ans\"atze
and by comparing with 
Eq.~\eqref{eqn:asympt-pot} up to $O(1/r^{2})$ corrections,
$F=dV/dr=\sigma-\gamma/r^{2}$.
We have found that the potential ansatz
including $\gamma/r$ describes the 
data well  and the force is 
consistently described 
by such a function, 
which have supported the universality of the $\gamma/r$
correction to the static potential.
Remarkably, we have also found that 
the function $F=\sigma-\gamma/r^{2}$
fits the force data down to relatively short distances
$r/r_{0}\sim 0.3$, which is in common with 
non-Abelian gauge theories.

\par
Secondly, we have measured the U(1) flux-tube profile 
(the electric field and the monopole current)
via the PLCF for the distances $r/r_{0}=0.217 - 2.35$.
We have investigated the width of the flux-tube 
profile as a function of~$r$ 
based on the dual Ginzburg-Landau (DGL) analysis; 
we have fitted the tail (large $\rho$ region) 
of the U(1) flux-tube profile
by the classical flux-tube solution of the DGL theory
and have extracted the dual gauge boson mass $m_{B}$,
whose inverse characterizes the width of the flux tube.
We have found that the mass is almost 
constant for larger~$r$ as $m_{B}r_{0}\sim 4.0$, which indicates that 
the width remains constant 
in the range $r/r_{0}=1.77 - 2.35$.
If we suppose that the width grows logarithmically
as predicted by the string model,
the question may be whether we can identify such 
a behavior within the limited range of~$r$.
This would require a further detailed investigation
especially for the range $r/r_{0} >1.77$.

\section*{Acknowledgment}

We are grateful to P. Weisz for constant encouragement 
and numerous discussions during the course of this work.
We are also indebted to M.~L\"uscher for critical discussions.
Y.K. wishes to thank E.-M.~Ilgenfritz, R.W.~Haymaker and T.~Matsuki 
for useful discussions.
M.K. is partially supported by Alexander von Humboldt
foundation, Germany.
P.M. was partially supported, at later stages, by Fonds zur F\"orderung
der wissenschaftlischen Forschung in \"Osterreich, project M767-N08.
The calculations were done on the NEC SX5 at 
Research Center for Nuclear Physics, 
Osaka University, Japan.

\clearpage
\appendix
\section{composite structure of the U(1) flux tube}
\label{sec:prof-decompose}

In this appendix, we show the 
numerical evidence of the 
composite structure of the U(1) flux tube.
This is possible by applying the 
Hodge decomposition to the external source to
identify its monopole and photon related parts 
and by measuring the corresponding profiles.
Unfortunately, since the 
Hodge decomposition requires 
the lattice Coulomb propagator
(see, Eq.~\eqref{eqn:def-ph-mo}), 
which spoils the locality of the operator, 
we cannot use the LW algorithm here.
Thus, we measure the flux-tube profile induced by 
the Wilson loop with a small size as an external source
instead of a PLCF to get a clear signal. 
How to decompose the Wilson loop into the monopole and 
photon parts is given below.
We explain this by using differential form notation.

\par
U(1) link variables $\theta (C_{1})$ can be decomposed into the 
electric-photon $\theta^{ph}(C_{1})$
and magnetic-monopole $\theta^{mo}(C_{1})$ parts 
in terms of $\bar{\theta}(C_{2})$ and $n(C_{2})$ as
\be
\theta = \Delta^{-1}\Delta \theta =  
\Delta^{-1} (d \delta + \delta d) \theta =
\Delta^{-1} d \delta \theta +
\Delta^{-1} \delta \bar{\theta}+ 2 \pi \Delta^{-1} \delta n \; ,
\label{eq:link_decomposition}
\ee
where we may define
\be
\theta^{ph} = \Delta^{-1} \delta \bar{\theta}\; , \qquad
\theta^{mo} = 2 \pi \Delta^{-1} \delta n \; .
\label{eqn:def-ph-mo}
\ee
To get the last equality 
of Eq.~\eqref{eq:link_decomposition}
we have used the relation 
$d\theta = \bar{\theta} + 2 \pi n$.
Note that the monopole part depends on $n$ while
the photon part does not.
The Wilson loop is then expressed as
\be
W_{A} \equiv 
\exp [i (\theta, j)] = \exp [i (\theta^{ph}, j)] 
\cdot \exp [i (\theta^{ph}, j)]
\equiv W_{\mathit{Ph}}\cdot W_{\mathit{Mo}} \; ,
\label{eq:wilsonloop_decomposition}
\ee
where the first term of the final expression 
in Eq.~\eqref{eq:link_decomposition}
does not contribute to the Wilson loop, since
we have
$(\Delta^{-1} d \delta \theta, j)
= (\Delta^{-1}  \delta \theta, \delta j) =0$
due to the conserved electric current $\delta j=0$.
In this sense, Eq.~\eqref{eq:wilsonloop_decomposition}
does not depend on the choice of the gauge.
It is known that the static potential extracted from 
$W_{\mathit{Ph}}$ and $W_{\mathit{Mo}}$
show a Coulombic and a linearly rising 
behaviors~\cite{Stack:1994ze}.

\par
In order to obtain the flux-tube profile, we
measure the correlation function
\bea
\langle \mathcal{O} \rangle_{j}
&=&
\frac{\langle  W \mathcal{O}\rangle_{0}}
{\langle W\rangle_{0}} \; ,
\label{eq:w-ope}
\eea
where $\mathcal{O}$ is a parity-odd local operator 
as Eqs.~\eqref{eq:ope-ele} and~\eqref{eq:ope-mono}.
Now, we may write $\mathcal{O} 
=\mathcal{O}_{\mathit{Ph}}+\mathcal{O}_{\mathit{Mo}}$.
We find that if relations
\be
\langle W \rangle_{0} \approx
\langle  W_{\mathit{Ph}}\rangle_{0}  
\langle W_{\mathit{Mo}} \rangle_{0}
\label{eq:require1}
\ee
and 
\be
\langle X_{\mathit{Ph}} 
Y_{\mathit{Mo}} \rangle_{0}\approx 0
\label{eq:require2}
\ee
are satisfied (where $X$ and $Y$ are arbitrary operators
but contain only the photon or monopole part), 
Eq.~\eqref{eq:w-ope} can further be evaluated as
\bea
\langle \mathcal{O}\rangle_{j}
&=&
\frac{\langle  (W_{\mathit{Ph}}\cdot W_{\mathit{Mo}})
(\mathcal{O}_{\mathit{Ph}}+\mathcal{O}_{\mathit{Mo}}) 
\rangle_{0}}{\langle W_{\mathit{Ph}}\cdot W_{\mathit{Mo}}\rangle_{0} }
=
\frac{\langle  W_{\mathit{Ph}}\mathcal{O}_{\mathit{Ph}} \rangle_{0}}
{\langle W_{\mathit{Ph}}\rangle_{0} }
+
\frac{\langle  W_{\mathit{Mo}}\mathcal{O}_{\mathit{Mo}} \rangle_{0}}
{\langle W_{\mathit{Mo}}\rangle_{0} } \nonumber\\*
&=&
\langle \mathcal{O}_{\mathit{Ph}} \rangle_{j}
+\langle \mathcal{O}_{\mathit{Mo}} \rangle_{j} \; .
\label{eq:expect-decompose}
\eea
This means that the sum of the profiles from the 
photon and monopole Wilson loops provides the full U(1) profile.
The validity of the assumptions, Eqs.~\eqref{eq:require1} 
and~\eqref{eq:require2}, will be checked in the data.

\par
\begin{figure}[t]
\includegraphics[height=5.5cm]{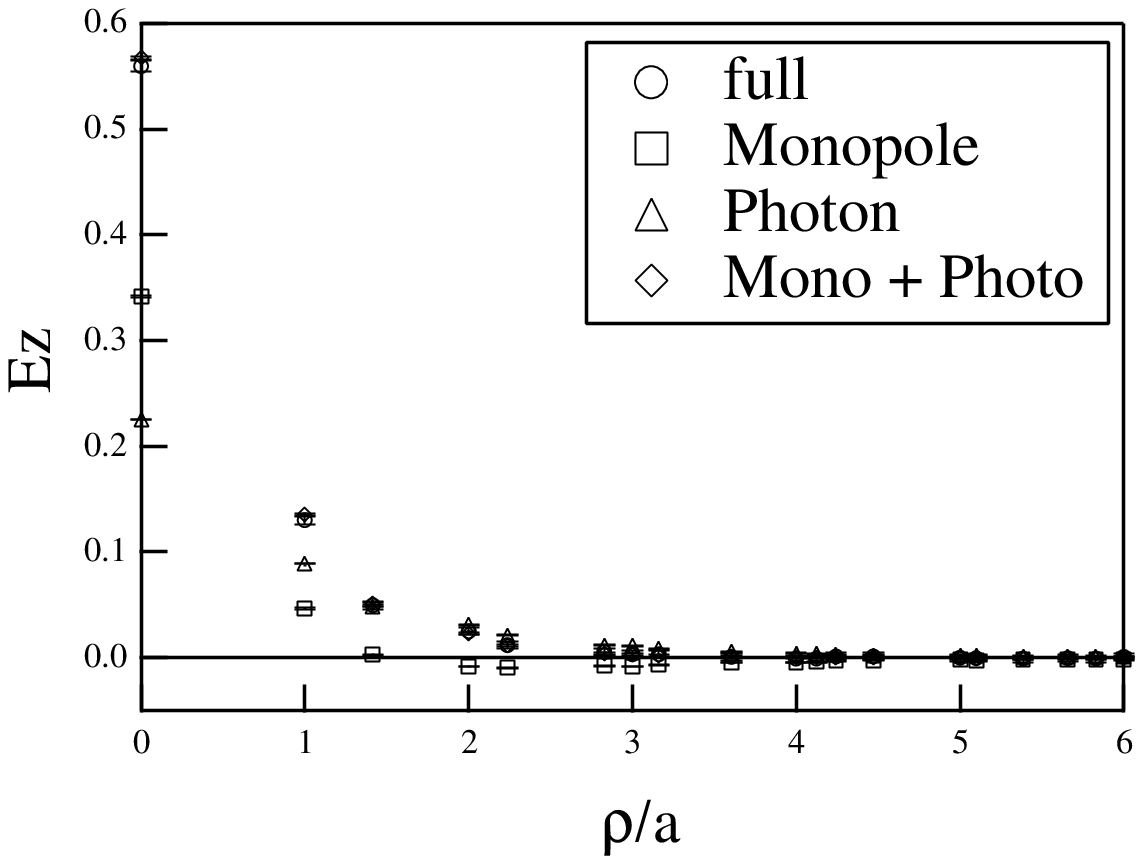}
\includegraphics[height=5.5cm]{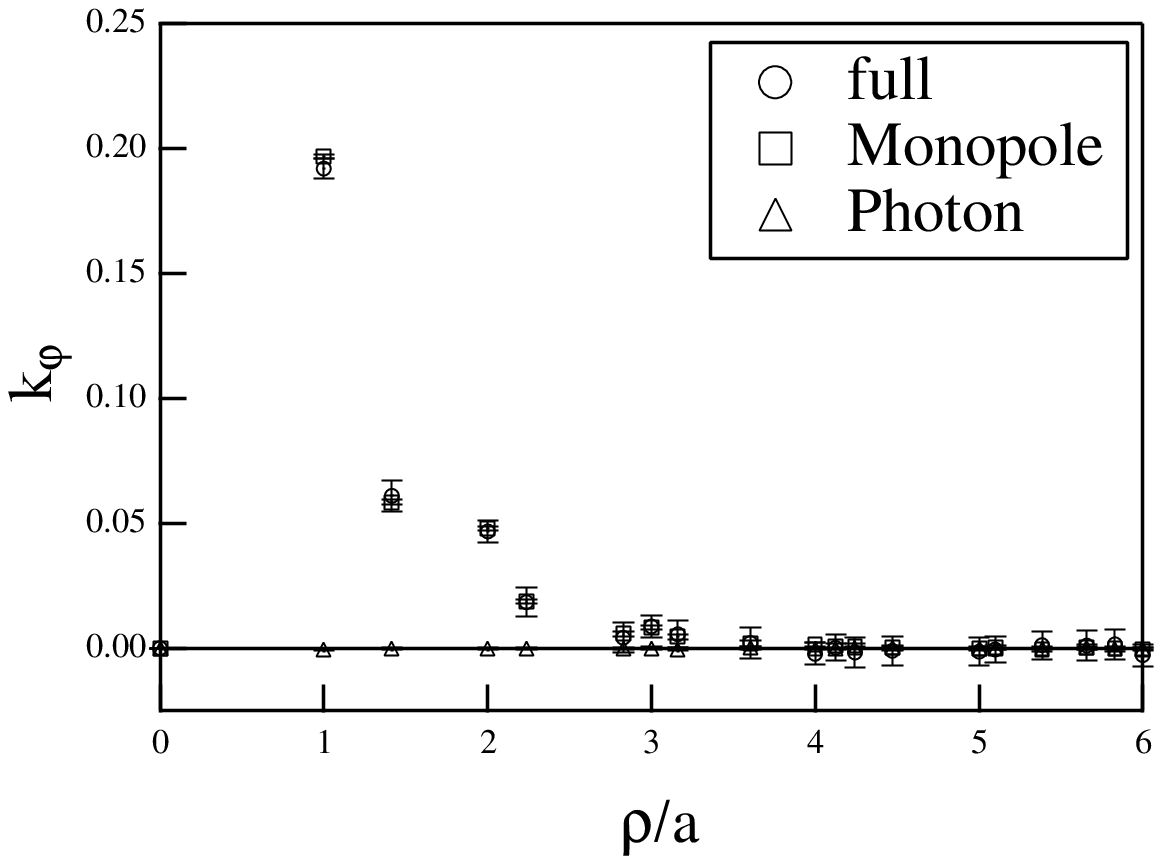}
\caption{Profiles of the electric field (left) and
the monopole current (right) from correlators with  U(1) 
Wilson loop ($3 \times 3$) and 
its photon and monopole parts.}
\label{fig:er_kr_33}
\end{figure}
\begin{figure}[t]
    \centering
\includegraphics[height=5.5cm]{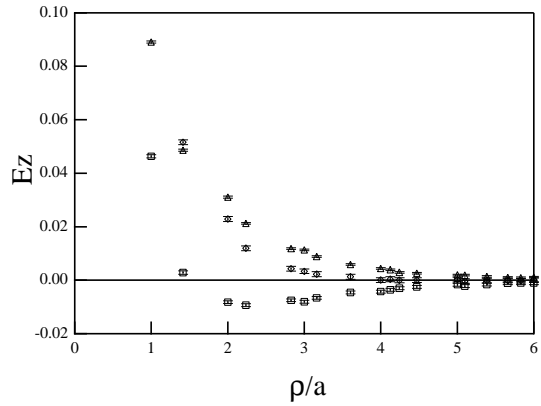}
\caption{The same plots as in Fig.~\ref{fig:er_kr_33}
with the $E_{z}$ axis rescaled.
The profile directly from the
full U(1) Wilson loop is omitted.}
\label{fig:er_kr_33_focus}
\end{figure}
\begin{figure}[t]
\centering
\includegraphics[width=10cm]{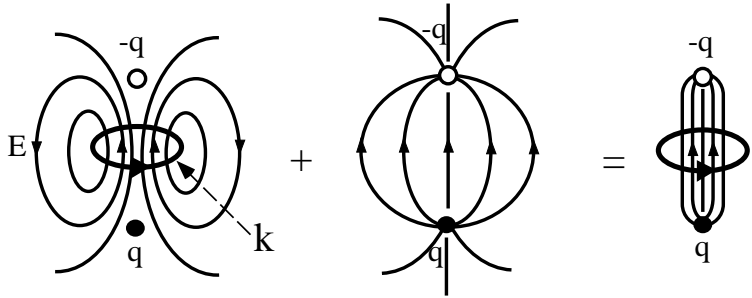}
\caption{The composite structure of the U(1) flux tube}
\label{fig:structure}
\end{figure}

In Fig.~\ref{fig:er_kr_33}, we show the
result with the $3\times 3$ Wilson loop at $\beta=0.99$.
Here we have used $N_{\rm conf}=1000$ configurations.
These profiles are measured just on the mid-point 
of the Wilson loop.
In Fig.~\ref{fig:er_kr_33_focus}, we
also show the same electric field profiles as in 
Fig.~\ref{fig:er_kr_33}, focussing
on the region around $E_{z} \sim 0$.
The important findings here are the following.
The full U(1) profile is given by the sum of the 
photon and monopole parts, which means that 
Eq.~\eqref{eq:expect-decompose} is satisfied.
The monopole part of $E_{z}$ becomes 
negative beyond a certain critical radius 
$\rho_{c}$ (in this case $\rho_{c}/a \sim 1.5$), 
indicating the appearance of the
solenoidal electric field, which 
cancels the Coulomb electric field from the
photon Wilson loop at $\rho > \rho_{c}$.
The corresponding schematic picture is  shown in 
Fig.~\ref{fig:structure}.
There is no correlation between the photon Wilson loop 
and monopole current and
only the monopole part is responsible for the 
monopole current profile.
This is consistent with the fact that the 
solenoidal electric field is from the 
monopole Wilson loop (dual Amp\`ere law).
Hence, we can conclude that the U(1) flux tube has the same 
composite structure as the classical flux-tube solution
of the DGL theory~\cite{Koma:2003gq}.
This result further supports the 
dual superconducting confinement mechanism 
(dual Meissner effect) in U(1) LGT.
The behavior of the profiles from  
$W_{\mathit{Ph}}$ and $W_{\mathit{Mo}}$
is completely consistent with that of the 
static potential~\cite{Stack:1994ze}.

\par
\begin{figure}[t]
\includegraphics[height=5.5cm]{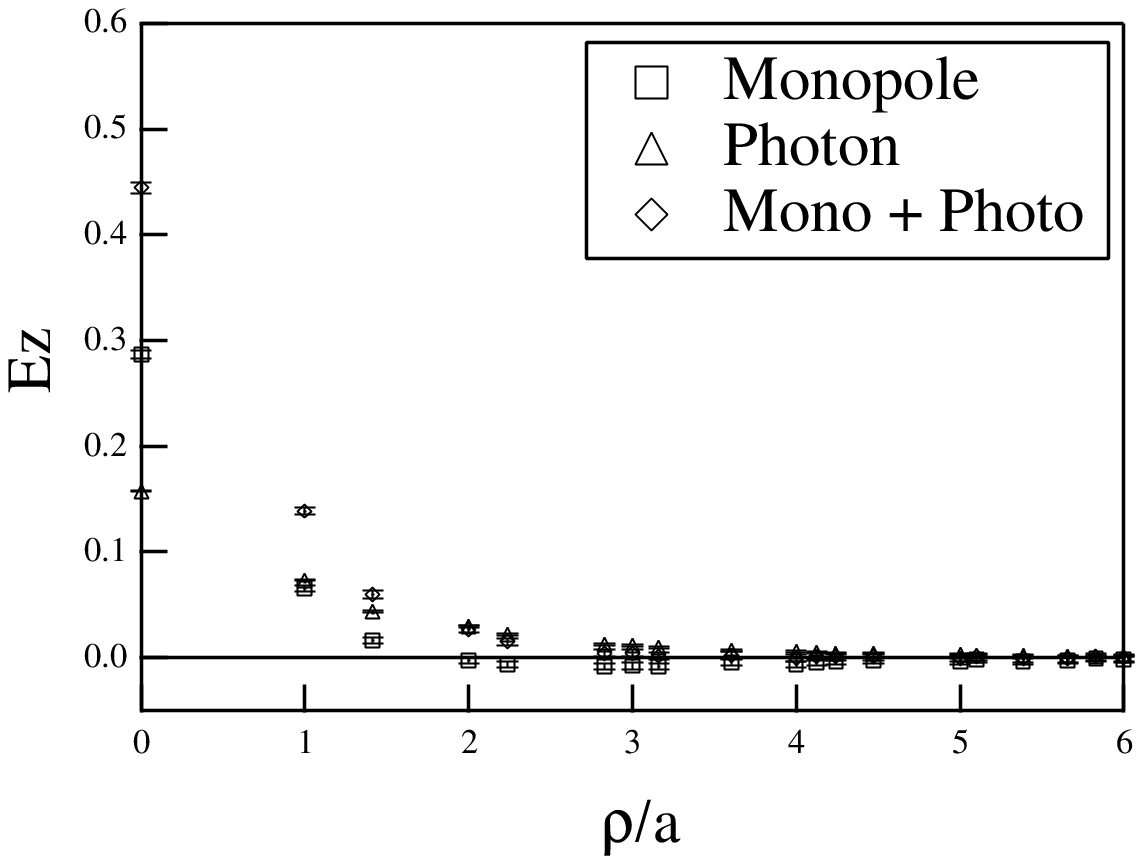}
\includegraphics[height=5.5cm]{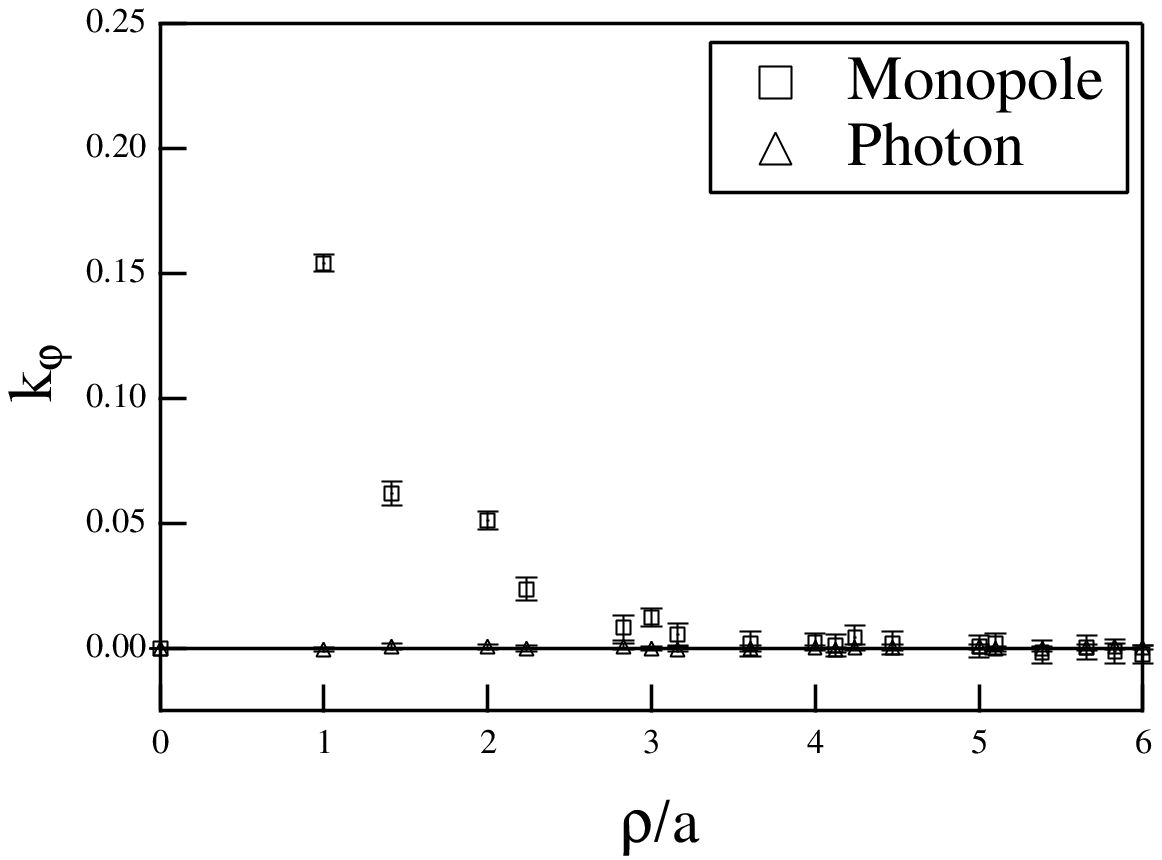}
\caption{The same plot as in Fig.~\ref{fig:er_kr_33} with W(3,5).
The profiles from the full U(1) Wilson loop are
omitted since they are too noisy.}
\label{fig:er_kr_35}
\end{figure}
\begin{figure}[t]
\includegraphics[height=5.5cm]{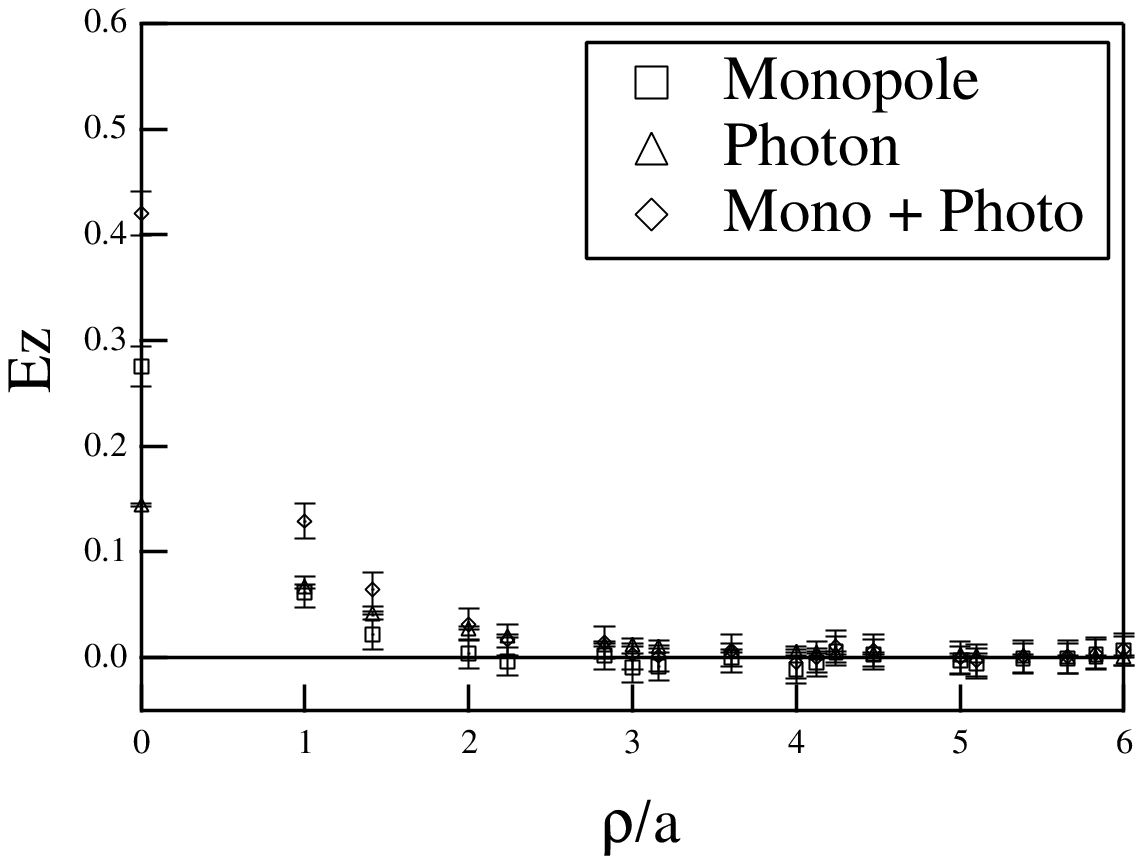}
\includegraphics[height=5.5cm]{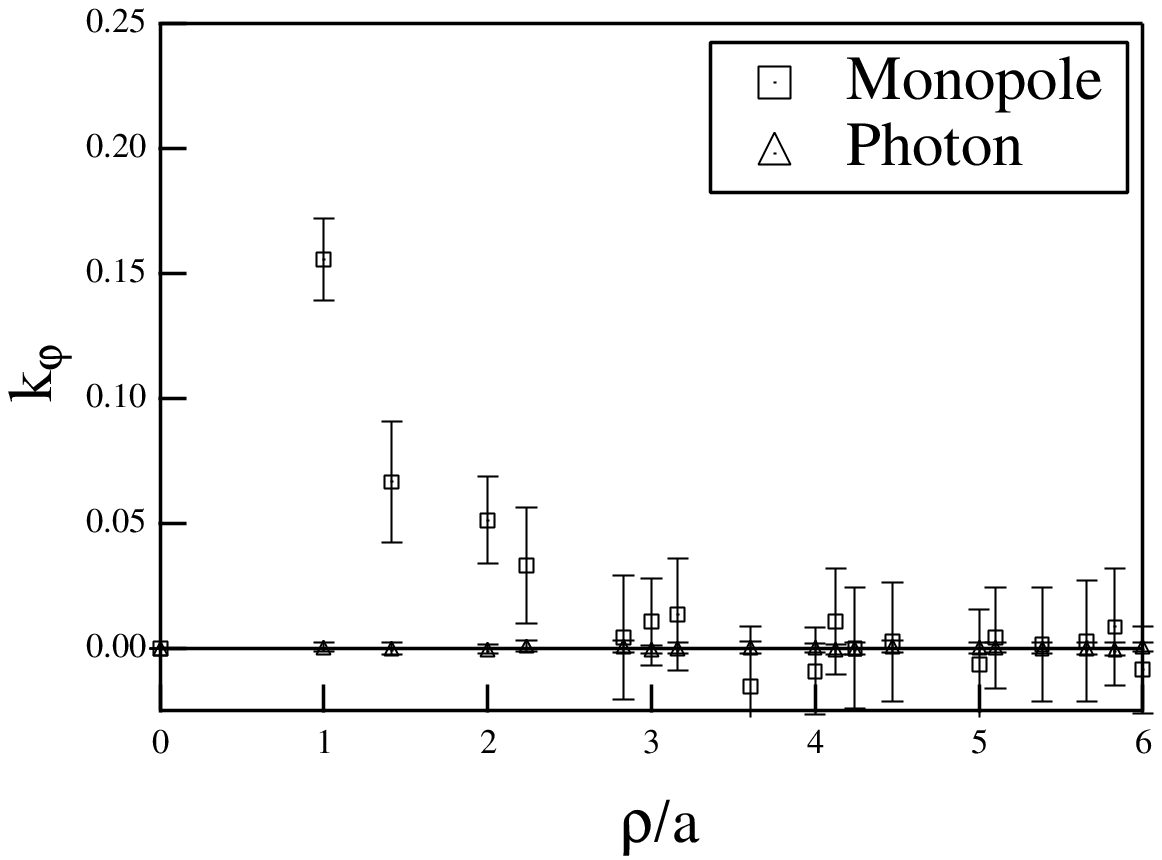}
\caption{The same plot as in Fig.~\ref{fig:er_kr_33}
with W(3,7).
The profiles from the full U(1) Wilson loop are omitted.}
\label{fig:er_kr_37}
\end{figure}

We find that the peak of the electric field 
is $E_{z}(0)=0.57$, which is larger than
that in Fig.~\ref{fig:b099-string}, $E_{z}(0)=0.38$,
obtained by using the PLCF with $r/a=3$.
This difference exhibits the effect of the finite
temporal extension $t$ of the Wilson loop;
the spatial part of the Wilson loop provides an 
additional Coulombic electric field.
In fact, as increasing $t$, we can observe that 
the profile approaches to the result from the PLCF, 
although it becomes difficult to find a clear signal 
to check such a behavior without the Hodge 
decomposition method of the Wilson loop
as shown in Figs.~\ref{fig:er_kr_35} 
and~\ref{fig:er_kr_37}.

\clearpage

\end{document}